\documentclass[fleqn,usenatbib]{mnras}

\usepackage{newtxtext,newtxmath}

\usepackage[T1]{fontenc}

\DeclareRobustCommand{\VAN}[3]{#2}
\let\VANthebibliography\thebibliography
\def\thebibliography{\DeclareRobustCommand{\VAN}[3]{##3}\VANthebibliography}

\usepackage{graphicx}	
\usepackage{amsmath}	
\usepackage{subcaption}
\usepackage{adjustbox}
\usepackage{xcolor}
\usepackage{soul}
\usepackage{svg}
\usepackage{ulem}
\usepackage{url}
\usepackage{xltabular}
\usepackage{aas_macros}
\usepackage{natbib}
\usepackage{longtable}

\urldef{\NEOCC}\url{https://neo.ssa.esa.int/search-for-asteroids?tab=summary&des=161989%20Cacus}

\title[Models of four NEAs]{Improved models for near-Earth asteroids (2100) Ra-Shalom, (3103) Eger, (12711) Tukmit \& (161989) Cacus}

\author[Javier Rodríguez Rodríguez et al.]{Javier Rodríguez Rodríguez,$^{1}$\thanks{E-mail: rodriguezrjavier@uniovi.es}
Enrique Díez Alonso,$^{1,2}$
Santiago Iglesias Álvarez,$^{1}$
Saúl Pérez Fernández,$^{1}$
\newauthor
Javier Licandro,$^{4,5}$\thanks{E-mail: jlicandr@iac.es}
Miguel R. Alarcon,$^{4,5}$
Miquel Serra-Ricart,$^{4,5}$
Noemi Pinilla-Alonso,$^{6}$
\newauthor
Susana Fernández Menéndez$^{1}$
and Francisco Javier de Cos Juez$^{1,3}$
\\ \\
$^{1}$Instituto Universitario de Ciencias y Tecnologías Espaciales de Asturias (ICTEA), University of Oviedo, C. Independencia 13, 33004 Oviedo, Spain\\
$^{2}$Departamento de Matemáticas, Facultad de Ciencias, Universidad de Oviedo, 33007 Oviedo, Spain\\
$^{3}$Departamento de Explotación y Prospección de Minas, Universidad de Oviedo, 33004 Oviedo, Spain\\
$^{4}$Instituto de Astrof\'{\i}sica de Canarias (IAC), C/V\'{\i}a L\'actea sn, 38205 La Laguna, Spain\\
$^{5}$Departamento de Astrof\'{\i}sica, Universidad de La Laguna, 38206 La Laguna, Tenerife, Spain\\
$^{6}$Florida Space Institute, University of Central Florida, Orlando, FL 32816, USA\\
}

\date{Accepted XXX. Received YYY; in original form ZZZ}

\pubyear{2023}

\begin{document}
\label{firstpage}
\pagerange{\pageref{firstpage}--\pageref{lastpage}}
\maketitle

\begin{abstract}

We present 24 new dense lightcurves of the near-Earth asteroids (3103) Eger, (161989) Cacus, (2100) Ra-Shalom and (12711) Tukmit, obtained with the Instituto Astrofísico Canarias 80 and Telescopio Abierto Remoto 2 telescopes at the Teide Observatory (Tenerife, Spain) during 2021 and 2022, in the framework of projects visible NEAs observations survey and NEO Rapid Observation, Characterization and Key Simulations. The shape models and rotation state parameters ($P$, $\lambda$, $\beta$) were computed by applying the lightcurve inversion method to the new data altogether with the archival data.
For (3013) Eger and (161989) Cacus, our shape models and rotation state parameters agree with previous works, though they have smaller uncertainties. For (2100) Ra-Shalom, our results also agree with previous studies. Still, we find that a Yarkovsky — O'Keefe — Radzievskii — Paddack acceleration of $\upsilon = (0.223\pm0.237)\times10^{-8}$ rad d$^{-2}$ slightly improves the fit of the lightcurves, suggesting that (2100) Ra-Shalom could be affected by this acceleration. We also present for the first time a shape model for (12711) Tukmit, along with its rotation state parameters ($P=3.484900 \pm 0.000031$ hr, $\lambda = 27^{\circ}\pm 8^{\circ}$, $\beta = 9^{\circ} \pm 15^{\circ}$).

\end{abstract}




\section{Introduction}

An asteroid is classified as a near-Earth asteroid (NEA) if it reaches its perihelion at a distance of less than 1.3 Astronomical Units (AU) from the Sun as stated in Center for Near Earth Object Studies (CNEOS)\footnote{\label{neo}\url{https://cneos.jpl.nasa.gov/about/neo_groups.html}} . Therefore, NEAs are the subgroup of minor bodies that come closest to the Earth. According to CNEOS\footnote{\label{neo2}\url{https://cneos.jpl.nasa.gov/stats/totals.html}}, as of 04/24/2023 there are 31,756 confirmed NEAs, of which 10,398 have a typical size greater than 140 m and 851 are larger than 1 km (the largest confirmed to date is (1036) Ganymed, with a diameter of $\sim$ 41 km, while the smaller known NEAs, as 2015 TC25, have radii of $\sim$ 1 m).

Among all the objects in this group, there is a subgroup known as Potentially Hazardous Asteroids (PHAs), which according to CNEOS\footref{neo}\footnote{\url{https://cneos.jpl.nasa.gov/glossary/PHA.html}} are those that represent a potential risk of collision with the Earth. More specifically, an asteroid is classified as PHA if its orbit has a Minimum Orbit Intersection Distance (MOID) with the Earth of 0.05 AU or less and its absolute magnitude is \textit{H} < 22, which implies that the object is larger than $\sim$ 140 m. These objects are fundamental due to their proximity to Earth and the possibility of a collision. By monitoring and studying these asteroids, we can accurately characterize and make them a potential resource source if their composition is rich in any interesting element. From the asteroids presented in this work (161989) Cacus, belongs to this group since its MOID is 0.014085 AU and its \textit{H} is 17.2 from data of European Space Agency (ESA) Near Earth Objects Coordination Centre (NEOCC)\footnote{\NEOCC}.

To obtain the models, it's widely applied the Convex Inversion Method detailed in \cite{2001Icar..153...24K,2001Icar..153...37K}, which generates a convex model and its corresponding spin state from a suitable set of lightcurves.
In the process, both the spin state and the shape are fitted at the same time, searching for the set of parameters (complete spin state and the corresponding shape) that best reproduce the observed lightcurves of the asteroid. The lightcurves can be \textit{dense} (that is, observations made at high cadence, of the order of minutes, and typically spanning a few hours) or \textit{sparse} (a few observations per night but typically extending for years). Dense lightcurves are usually the result of specific follow-up programs, such as the Visible NEAs Observations Survey (ViNOS; \cite{2023MNRAS.tmp..693L}), while sparse lightcurves are usually obtained from surveys that periodically patrol the sky such as the Asteroid Terrestrial-impact Last Alert System (ATLAS; \cite{2018AJ....156..241H,2018PASP..130f4505T}), the All-Sky Automated Survey for Supernovae (ASAS-SN; \cite{2017PASP..129j4502K}) or the Wide Angle Search for Planets (SuperWASP; \cite{2005EM&P...97..261P}) among many others. In the lightcurve inversion process, it's possible to work only with dense data \citep{2003Icar..164..346T, 2007A&A...465..331D}, only sparse data \citep{2016A&A...587A..48D,2019A&A...631A...2D} or a well-balanced combination of both \citep{2009A&A...493..291D}. However, to obtain reliable results, the lightcurves must be acquired by covering the widest possible range of phase angles, which results in observations corresponding to different geometries that encode information related to the main features of the asteroids. A large number of asteroid models, along with their parameters, lightcurves and many other products, is available at the Database of Asteroid Models from Inversion Techniques (DAMIT\footnote{\url{https://astro.troja.mff.cuni.cz/projects/damit/}}; \cite{2010A&A...513A..46D}), operated by The Astronomical Institute of the Charles University (Prague, Czech Republic).

 Small asteroids make up the vast majority of the NEA population (97.3\% is estimated to have a diameter smaller than 1 km, according to CNEOS\footref{neo2}). Two critical mechanisms acting on these small bodies are the Yarkovsky \citep{yarkovsky1901density,2006AREPS..34..157B,2015aste.book..509V} and the Yarkovsky-O'Keefe-Radzievskii-Paddack (YORP; \cite{yarkovsky1901density,1952AZh....29..162R,1969JGR....74.4379P,o1976tektites,2006AREPS..34..157B,2015aste.book..509V}) effects. The first consists of orbital changes due to thermal reemision of the absorbed solar radiation, increasing the orbit's semi-major axis  if the asteroid is a prograde rotator and decreasing it otherwise. It also plays a crucial role in injecting new NEAs from the Main Asteroid Belt \citep{2003Sci...302.1739C,2003Icar..163..120M}. The YORP effect is a  constant change in the spin state caused by anisotropic thermal re-emission and the resulting torque.
 
 There are several observations attributed to the YORP effect that are considered as indirect detections. One is the clustering in the directions of the rotation axes among members of the same asteroid family; for example, this clustering has been observed among the Koronis members  \citep{2002Natur.419...49S}. It is also thought to be responsible of the bimodalities observed in the rotation rates \citep{2008Icar..197..497P} and obliquities \citep{2013A&A...559A.134H} for small asteroids. Furthermore, it is believed to be a prominent mechanism in the formation of small binaries \citep{2008Natur.454..188W}.
 
The first direct detection of the YORP effect was in the NEA (6489) Golevka utilizing radar techniques \citep{2003Sci...302.1739C}. Later it has also been detected from photometric data in (1862) Apollo \citep{2007Natur.446..420K}, (54509) 2000 PH5 \citep{2007Sci...316..272L, 2007Sci...316..274T}, (1620) Geographos \citep{2008A&A...489L..25D}, (25143) Itokawa, \citep{2014A&A...562A..48L}, (1685) Toro, (3103) Eger and (161989) Cacus \citep{2018A&A...609A..86D}.

In Section \ref{sec:Observations} of this work, we present new dense lighcurves of the NEAs (2100) Ra-Shalom, (3103) Eger, (12711) Tukmit and (161989) Cacus, acquired at Teide Observatory. In Section \ref{sec:Methods} we explain how these observations have been processed along with archival lightcurves to compute the shape models and rotational state applying the lightcurve inversion method. Results are presented and compared to previous published models in Section \ref{sec:Discussion}. Finally, our conclusions are presented in Section \ref{sec:Conclussions}.

\section{Observations}
\label{sec:Observations}

Time series photometry of NEAs (2100) Ra-Shalom, (3103) Eger, (12711) Tukmit and (161989) Cacus were obtained in the framework of ViNOS  \citep{2023MNRAS.tmp..693L}, aimed to characterize NEAs by using spectroscopic, spectro-photometric, and lightcurves observations, and the NEO Rapid Observation, Characterization and Key Simulations (NEOROCKS\footnote{\url{https://www.neorocks.eu/}}) project, where the Instituto Astrofísico Canarias (IAC) team lead the task on the characterization of radar targets. We note that the NEAs studied in this paper were observed using radar: 2100 in \citet{1984Icar...60..391O} and \citet{2000Icar..147..520S,2008Icar..195..184S}; 3103 in \citet{1997Icar..130..296B}, 12711 in \citet{2008Icar..198..294B}; and 161989 with Goldstone in 2022 August 24\footnote{\url{https://echo.jpl.nasa.gov/asteroids/Cacu/Cacus.2022.goldstone.planning.html}}.

 Photometric observations were obtained using two telescopes located at Teide Observatory (TO, Tenerife, Canary Islands, Spain), the Instituto Astrofísico Canarias 80 (IAC80) and Telescopio Abierto Remoto 2 (TAR2) telescopes. The observational circumstances are shown in Table \ref{tab_obs_cir}. 

\begin{table*}
\caption{Observational circumstances of new lightcurves acquired by ViNOS. The table includes the object, telescope and filters used (r-sloan, V, Clear and Luminance), the date and the starting and end time (UT) of the observations, the phase angle ($\alpha$), the heliocentric ($r$) and geocentric ($\Delta$) distances and phase angle bisector longitude (PABLon) and latitude (PABLat) of the asteroid at the time of observation.}          
\label{tab_obs_cir} 
\resizebox{\textwidth}{!}{%
\begin{tabular}{llcclllrrrrr}
\hline
 Asteroid                 & Telescope   & Filter   &   Exp. Time [s] & Date        & UT (start)   & UT (end)    &   $\alpha [^\circ]$ &   $r$ [au] &   $\Delta$ [au] &   PABLon [deg] &          PABLat [deg] \\
\hline
 2100 Ra-Shalom (1978 RA) & IAC80        & r        &           45 & 2022-Jul-29 & 00:45:17.539 & 5:24:43.286 &               68.35 &     1.0858 &          0.2885 &       349.56   &        25.5341 \\
 2100 Ra-Shalom (1978 RA) & IAC80        & r        &           45 & 2022-Aug-02 & 00:34:00.941 & 5:12:47.808 &               64.28 &     1.1031 &          0.2725 &       349.861  &        24.4537 \\
 2100 Ra-Shalom (1978 RA) & TAR2        & L        &           60 & 2022-Aug-05 & 01:32:32.755 & 5:07:01.430 &               60.92 &     1.1152 &          0.2602 &       349.888  &        23.6158 \\
 2100 Ra-Shalom (1978 RA) & TAR2        & L        &           30 & 2022-Aug-24 & 20:58:08.803 & 0:52:10.186 &               30.6  &     1.1731 &          0.194  &       344.45   &        16.4497 \\
 2100 Ra-Shalom (1978 RA) & TAR2        & L        &           30 & 2022-Aug-26 & 20:33:01.210 & 3:59:47.558 &               26.93 &     1.1769 &          0.1907 &       343.353  &        15.4405 \\
 2100 Ra-Shalom (1978 RA) & TAR2        & L        &           30 & 2022-Sep-06 & 20:17:37.248 & 1:17:07.642 &               16.42 &     1.1915 &          0.1929 &       336.953  &         8.9531 \\
 2100 Ra-Shalom (1978 RA) & TAR2        & L        &           30 & 2022-Sep-08 & 00:19:01.430 & 3:13:47.453 &               17.28 &     1.1924 &          0.1953 &       336.353  &         8.2298 \\
 3103 Eger (1982 BB)      & TAR2        & Clear        &           60 & 2021-Jul-03 & 00:51:30.010 & 4:29:21.984 &               51.95 &     1.2062 &          0.3806 &       323.062  &        10.7387 \\
 3103 Eger (1982 BB)      & TAR2        & Clear        &           60 & 2021-Jul-04 & 00:45:43.978 & 5:14:49.027 &               52.32 &     1.2007 &          0.3723 &       323.984  &        10.2764 \\
 3103 Eger (1982 BB)      & TAR2        & Clear        &           60 & 2021-Jul-05 & 00:46:34.003 & 5:14:39.005 &               52.71 &     1.1952 &          0.364  &       324.927  &         9.7906 \\
 3103 Eger (1982 BB)      & TAR2        & Clear        &           50 & 2021-Jul-17 & 01:46:36.019 & 5:18:16.992 &               59.39 &     1.1299 &          0.2796 &       337.787  &         1.8092 \\
 3103 Eger (1982 BB)      & TAR2        & Clear       &           50 & 2021-Jul-18 & 01:46:36.970 & 5:27:30.038 &               60.16 &     1.1246 &          0.2742 &       338.986  &         0.9327 \\
 3103 Eger (1982 BB)      & TAR2        & Clear        &           50 & 2021-Jul-19 & 01:46:46.042 & 5:30:09.965 &               60.96 &     1.1193 &          0.2691 &       340.204  &         0.0191 \\
 3103 Eger (1982 BB)      & TAR2        & V        &           90 & 2021-Dec-13 & 02:18:08.957 & 6:13:05.030 &               54.13 &     1.213  &          0.653  &       140.013  &       -10.1599 \\
 3103 Eger (1982 BB)      & TAR2        & V        &           60 & 2022-Feb-12 & 01:54:58.954 & 6:55:11.021 &               12.56 &     1.5281 &          0.562  &       149.456  &        11.7599 \\
 3103 Eger (1982 BB)      & TAR2        & V        &           60 & 2022-Feb-13 & 01:21:01.037 & 6:53:21.034 &               12.3  &     1.5326 &          0.5656 &       149.371  &        12.022  \\
 3103 Eger (1982 BB)      & TAR2        & V        &           90 & 2022-Mar-01 & 20:04:46.992 & 2:10:05.030 &               17.04 &     1.6053 &          0.6627 &       148.428  &        15.5233 \\
 12711 Tukmit (1991 BB)   & TAR2        & V        &           90 & 2021-Dec-28 & 02:57:19.469 & 6:45:29.030 &               26.94 &     1.433  &          0.5388 &       123.23   &         7.243  \\
 12711 Tukmit (1991 BB)   & TAR2        & V        &           90 & 2022-Aug-04 & 21:00:01.037 & 0:05:17.261 &               89.3  &     0.9779 &          0.2827 &       333.589  &        58.4229 \\
 12711 Tukmit (1991 BB)   & TAR2        & V        &           60 & 2022-Sep-05 & 20:29:33.590 & 3:52:59.750 &               64.07 &     1.1164 &          0.5806 &       347.955  &        63.1137 \\
 161989 Cacus (1978 CA)   & IAC80        & r        &           20 & 2022-Feb-22 & 20:09:12.154 & 3:56:50.352 &               45.52 &     1.2199 &          0.3846 &       121.14   &       -22.2578 \\
 161989 Cacus (1978 CA)   & TAR2        & L        &           20 & 2022-Aug-25 & 01:00:10.310 & 1:43:30.518 &               93.77 &     1.0022 &          0.0825 &        15.7717 &        29.7916 \\
 161989 Cacus (1978 CA)   & TAR2        & L        &           10 & 2022-Sep-04 & 01:30:17.885 & 5:38:49.229 &               61.49 &     1.0367 &          0.0619 &        12.2694 &       -14.1511 \\
\hline
\end{tabular}%
}

\end{table*}

The IAC80 is a 82~cm telescope with  $f/D =$ 11.3 in the Cassegrain focus. It is equipped with the CAMELOT-2 camera, a back-illuminated e2v 4K x 4K pixels CCD of 15 \textmu m$^2$ pixels, a plate scale of 0.32 arcsec/pixel, and a field of view of 21.98 x 22.06 arcmin$^2$. We used a Sloan $r$ filter. Observations were done using sidereal tracking, so the asteroid's proper motion limited the images' individual exposure time. We selected exposure times such that the asteroid trail was smaller than the typical FWHM of the IAC80 images ($\sim 1.0$ \arcsec). The images were bias and flat-field corrected in the standard way; there was not needed to correct the dark current since it is almost 0 for these CCD, so correcting the bias is enough. 

TAR2 is a 46~cm $f/D =$ 2.8 robotic telescope. Until July 2022 TAR2 was equipped with a FLI-Kepler KL400 camera, since then is equipped with a QHY600PRO camera.  The FLI-Kepler KL400 camera has a back illuminated 2K x 2K pixels GPixel GSense400 CMOS with a pixel size of 11 \textmu m$^2$ that in the prime focus of TAR2 has a plate scale of 1.77 arcsec/pixel and a field of view of $\sim$ 1 $\text{deg}^2$. The QHY600PRO camera detector is a Sony back illuminated 9K x 6K pixels IMX455 CMOS of 3.76 \textmu m$^2$ pixels, that in the prime focus of TAR2 has a plate scale of 0.65 arcsec/pixel and a field of view of $\sim$ 1.6 x  1.1 $\text{deg}^2$. Both CMOS use a rolling shutter and have the advantage of zero dead-time between images. For a complete description of the QHY600PRO capabilities see \cite{2023arXiv230203700A}. The images were bias, dark and flat-field corrected in the standard way. With both cameras we obtained a continuous series of 10 seconds images without filter (Clear) or using a Johnson $V$ filter with the FLI camera and a UV/IR cut L-filter with the QHY with the telescope moving in sidereal tracking. To increase the SNR, consecutive images were aligned and combined to produce a final series of images of larger exposure time. In general, the number of images used to obtain the final combined one is determined by the proper motion of the NEA. This is computed such that the total exposure time is shorter than the time it takes for the asteroid trail to be equal to the typical FWHM of this telescope ($\sim$ 3.6\arcsec). 

To obtain the lightcurves, we did aperture photometry of the final images using the Photometry Pipeline\footnote{\url{https://photometrypipeline.readthedocs.io/en/latest/}} (PP) software \citep{2017A&C....18...47M}, as we did in \citep{2023MNRAS.tmp..693L}. The images obtained with the L-filter were calibrated to the $r$ SLOAN band using the Pan-STARRS catalogue while the other images were calibrated to the corresponding bands for the filters used.

The new lightcurves are presented in Appendix \ref{sec: Fits of models and data} along with the synthetic models computed following the method explained in Section \ref{sec:Methods} (see Figures \ref{fig:IAC fit Ra-Shalom} for (2100) Ra-Shalom, \ref{fig:IAC fit Eger} for (3103) Eger, \ref{fig:IAC fit Cacus} for (161989) Cacus, and Figure \ref{fig:Fit Tukmit} in Section \ref{subsec:Tukmit} for (12711) Tukmit).


\section{Methods}
\label{sec:Methods}

When discussing asteroid characterization, some basic parameters are needed to create the asteroid's model that we further describe next. First of all, the sidereal rotation period ($P$), is the time the asteroid takes to complete a single revolution over its rotation axis and adopt the background stars as the reference frame. It is derived from the asteroid lightcurves applying periodogram-type tools. Lambda ($\lambda$) and Beta ($\beta$) are the ecliptic coordinates towards which the spin axis of the asteroid points, being $\lambda$ the ecliptic longitude ($0^\circ < \lambda \leq 360^\circ$), and $\beta$ the ecliptic latitude ($-90^\circ \leq \beta \leq 90^\circ$). With the pole solution ($\lambda$, $\beta$) and the asteroid's inclination (i), longitude of ascending node ($\Omega$) and the argument of pericenter ($\omega$), the obliquity ($\epsilon$) is then obtained. In the case of $0^\circ \leq \epsilon \leq 90^\circ$ , the asteroid will have a prograde rotation and retrograde otherwise ($90^\circ < \epsilon \leq 180^\circ$). It is possible to obtain a pole ambiguity for $\lambda$, that is, we could obtain two solutions with almost the same value for $\beta$, and a pair of values for $\lambda$ that differ $\sim$ 180$^{\circ}$ between each other.

In this work, we used our new lightcurves presented in Section \ref{sec:Observations}, along with available sets of archival lightcurves. All the archival lightcurves were obtained from the DAMIT and Asteroid Lightcurve Data Exchange Format (ALCDEF; \cite{2018DPS....5041703S}) databases. In Tables \ref{tab:Ra-Shalom archive}, \ref{tab:Eger archive}, \ref{tab:Tukmit archive} and \ref{tab:Cacus archive} we summarize the archival lightcurves used for each asteroid.

We applied the lightcurve inversion method to the set of lightcurves for each asteroid with two codes. The first one (No YORP code) was utilized. e.g., in \cite{2010A&A...513A..46D} or \cite{2011A&A...530A.134H}. It generates models with constant $P$ and is publicly available at the DAMIT website. The second code used (YORP code) is a modification of the former, which allows for linear evolution in $P$ over time, thus allowing to detect if the asteroid exhibits the YORP effect. It was gently provided by Josef \v{D}urech in personal communication; since it is not publicly available, the code was used in previous studies as \citet{2012A&A...547A..10D}.

For each asteroid we applied the following procedure independently with the No YORP and the YORP codes; Firstly, we obtained a medium resolution solution searching for $\lambda$ and $\beta$ values in all the sphere ($0^{\circ} < \lambda \leq 360^{\circ}, -90^{\circ} \leq \beta \leq 90^{\circ}$) with $5^{\circ}$ steps and adopting as initial value for $P$ the previously accepted value (except for (12711) Tukmit, for which we used the $P$ found with the period search tool implemented in the DAMIT code). Secondly, we performed a fine pole search with $2^{\circ}$ steps in a $30^{\circ}$ x $30^{\circ}$ square centered on the previous solution and starting with the $P$ obtained in the previous search. The initial parameters for modelling were set to their default (and recommended) values; in the case of the YORP code, the YORP value was set to $\upsilon=1\times10^{-8}$. Only the convexity regularization weight was modified in order to maintain the dark facet area below 1\% when needed. After running both codes, we reduce the solution's $\chi^2$ given by the code, to the number of measurements for each asteroid, obtaining a $\chi_\mathrm{red}^{2}$ value, selecting as a final solution the one with the lowest $\chi_\mathrm{red}^{2}$ value.

To obtain the uncertainties of the solution we opted for creating 100 subsets from the main set of measurements that was used to obtain the best-fitting solution in terms of $\chi_\mathrm{red}^{2}$. To create this subsets, we removed randomly 10\% or 25\% of the measurements from the initial set depending on its measurement number. We then recalculated the best-fitting solution for each of this new subsets, repeating the fine pole search, thus obtaining 100 solutions. With this 100 solutions, we then calculated the mean (which is almost identical to the best-fitting solution using the initial set of measurements) and standard deviation (3$\sigma$ level) which are the uncertainty of the solution.

Furthermore we applied the method proposed in \citet{2017AJ....153..270V} to alternatively obtain the uncertainty in the YORP effect at the $3\sigma$ level. For that we iterated the YORP code with all parameters, besides the YORP effect, fixed at the initial best-fitting solution values, modifying only the $\upsilon$ parameter and finally adopting as the final solution the one corresponding to the lowest $\chi_\mathrm{red}^{2}$ value (see Figure \ref{fig:Eger Yorp range} as an example).

\section{Results and Discussion}
\label{sec:Discussion}

We proceed now to show the results obtained following the methods proposed in Section \ref{sec:Methods} with a discussion for each asteroid (see Table \ref{tab_results} for a summary of the values obtained).

\begin{table*}
\caption{Results obtained in this work for each asteroid, we show type of model (linearly increasing period (L) and constant period (C)), rotation period, geocentric ecliptic coordinates of the spin pole ($\lambda, \beta$), obliquity ($\epsilon$) and YORP acceleration ($\upsilon$) if the model has linearly increasing period (L).}
\label{tab_results} 
\begin{tabular}{llrrrrr}
\hline
Asteroid                 & Model & Period [hr]          & $\lambda [^\circ]$           & $\beta [^\circ]$            & $\epsilon [^\circ]$          & $\upsilon$ [rad d$^{-2}$] \\
\hline
2100 Ra-Shalom (1978 RA) & C     & 19.820056$\pm$0.000012 & 278$\pm$18         & -60$\pm$8        & 162$\pm$10          & -                              \\
2100 Ra-Shalom (1978 RA) & L     & 19.820107$\pm$0.000040 & 278$\pm$8          & -60$\pm$5        & 165$\pm$5           & (0.22$\pm$0.16)$\times10^{-8}$ \\
3103 Eger (1982 BB)      & L     & 5.710148$\pm$0.000006  & 214$\pm$3          & -71$\pm$1        & 177$\pm$1           & (0.85$\pm$0.05)$\times10^{-8}$ \\
12711 Tukmit (1991 BB)   & C     & 3.484900$\pm$0.000031  & 27$\pm$8           & 9$\pm$15         & 119$\pm$15          & -                              \\
161989 Cacus (1978 CA)   & L     & 3.755067$\pm$0.000001  & 251$\pm$6          & -62$\pm$2        & 177$\pm$2           & (1.91$\pm$0.05)$\times10^{-8}$ \\
\hline
\end{tabular}
\end{table*}

\subsection{(2100) Ra-Shalom}
\label{subsec:Ra-Shalom}
 In previous studies \citep{2004Icar..167..178K,2012A&A...547A..10D,2018A&A...609A..86D} a rotation state parameters of $P=19.8200 \pm 0.0003$ hr, $\lambda = 295^{\circ} \pm 15^{\circ}$ and $\beta = -65^{\circ} \pm 10^{\circ}$ were reported as the most probable solution, and no YORP effect was detected. In these previous works 105 lightcurves from 
 \cite{1984Icar...60..391O},
 \cite{1992Icar...95..115H}, \cite{1998Icar..136..124P},
\cite{2004Icar..167..178K} and
\cite{2012A&A...547A..10D,2018A&A...609A..86D} were used, spanning from 1978 to 2016.

\begin{figure}
    \centering
    \includegraphics[width=\linewidth,keepaspectratio]{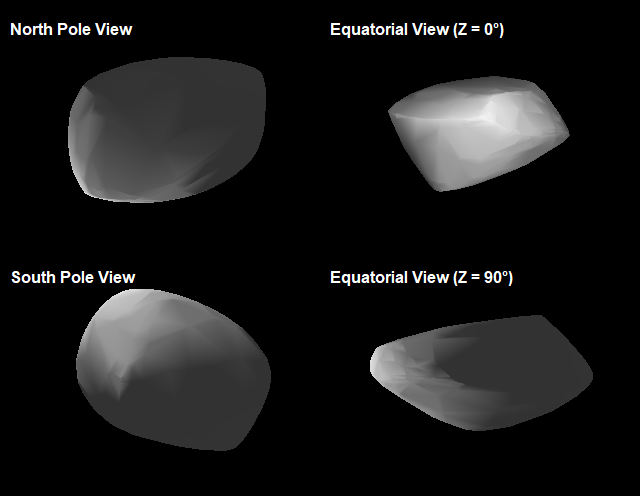}
    \caption{Constant rotation period shape model of (2100) Ra-Shalom. Left top: North Pole View (Y axis = 0$^{\circ}$). Left bottom: South Pole View (Y axis = 180$^{\circ}$). Right top and bottom: Equatorial Views with Z axis rotated 0$^{\circ}$ and 90$^{\circ}$.}
    \label{fig:Ra-Shalom No Yorp model}
\end{figure}

We applied the inversion algorithm to 93 archival lightcurves and our 7 new lightcurves acquired during 2022 (see Tables \ref{tab_obs_cir} and \ref{tab:Ra-Shalom archive}). First of all we ran the No YORP code since no linear evolution of $P$ was previously reported. Figure \ref{fig:Ra-Shalom No Yorp model} shows the shape model obtained with this code, corresponding to a pole solution $\lambda = 278^{\circ}$, $\beta = -60^{\circ}$, $\epsilon \simeq164^{\circ}$ and a rotation period of $P = 19.820056$ hr. The fit between the model and the data results in $\chi_\mathrm{red}^{2}=1.66$ normalized to the 4987 data points (See Figure \ref{fig:Ra-Shalom pole plot no yorp}).

\begin{figure}
    \centering
    \includegraphics[width=\linewidth,keepaspectratio]{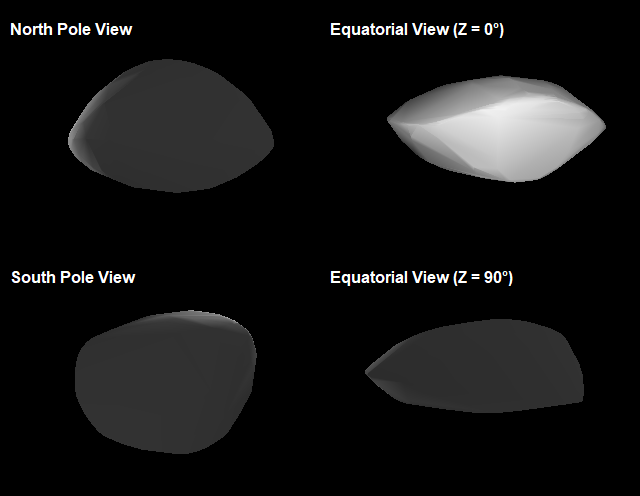}
    \caption{Linearly increasing rotation period shape model of (2100) Ra-Shalom. Left top: North Pole View (Y axis = 0$^{\circ}$). Left bottom: South Pole View (Y axis = 180$^{\circ}$). Right top and bottom: Equatorial Views with Z axis rotated 0$^{\circ}$ and 90$^{\circ}$.}
    \label{fig:Ra-Shalom Yorp model}
\end{figure}

\begin{figure*}
    \centering
    \begin{subfigure}{0.49\textwidth}
        \centering
        \includegraphics[width=\textwidth]{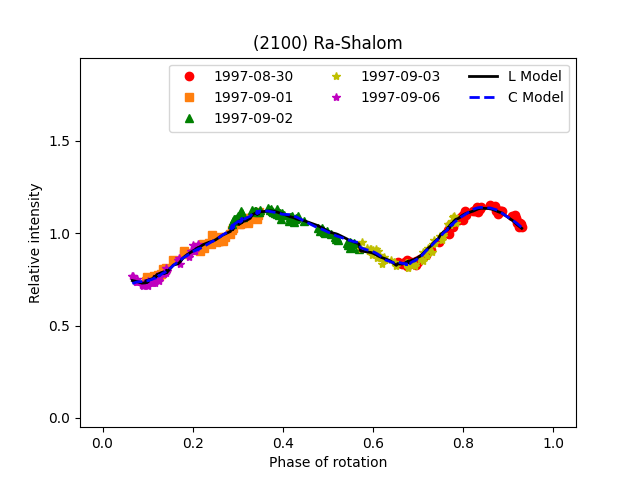}
    \end{subfigure}
    \begin{subfigure}{0.49\textwidth}
        \centering
        \includegraphics[width=\textwidth]{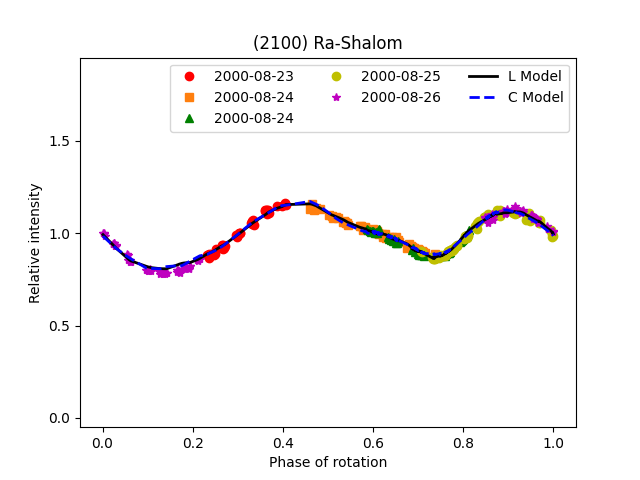}
    \end{subfigure}
    \begin{subfigure}{0.49\textwidth}
        \centering
        \includegraphics[width=\textwidth]{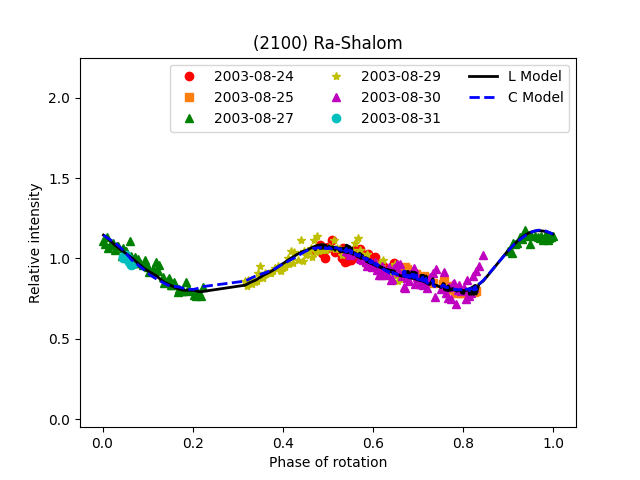}
    \end{subfigure}
    \begin{subfigure}{0.49\textwidth}
        \centering
        \includegraphics[width=\textwidth]{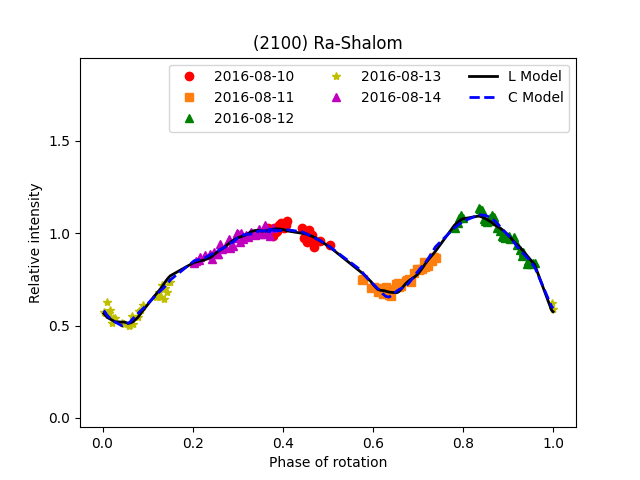}
    \end{subfigure}
    \caption{Fits between sets of lightcurves of (2100) Ra-Shalom corresponding to the 1997, 2000, 2003 \& 2016 seasons and the best-fitting models. Dashed blue: best constant period model (C Model). Solid black: best linearly increasing period model (L Model). Data for each observation represented by the colour and shapes shown in each legend.}
    \label{fig:Fit Ra-Shalom}
\end{figure*}

Next we performed the inversion with the YORP code, obtaining the shape model presented in Figure \ref{fig:Ra-Shalom Yorp model}, with the pole solution $\lambda = 283^{\circ}$, $\beta = -62^{\circ}$,  $\epsilon \simeq165^{\circ}$ a rotation period of $P = 19.820101$ hr (corresponding to 12 September 1978) and a YORP acceleration $\upsilon = 0.19\times10^{-8}$ rad d$^{-2}$. In this case the fit between the model and the data was slightly better, resulting in $\chi_\mathrm{red}^{2}=1.64$ normalized to the 4987 data points (See Figure \ref{fig:Ra-Shalom pole plot}). In Figure \ref{fig:Fit Ra-Shalom} we show the fits between the constant period (No YORP) and linearly increasing period (YORP) models for Ra-Shalom and the data corresponding to several seasons of observations.

\begin{figure}
    \centering
    \includegraphics[width=\linewidth,keepaspectratio]{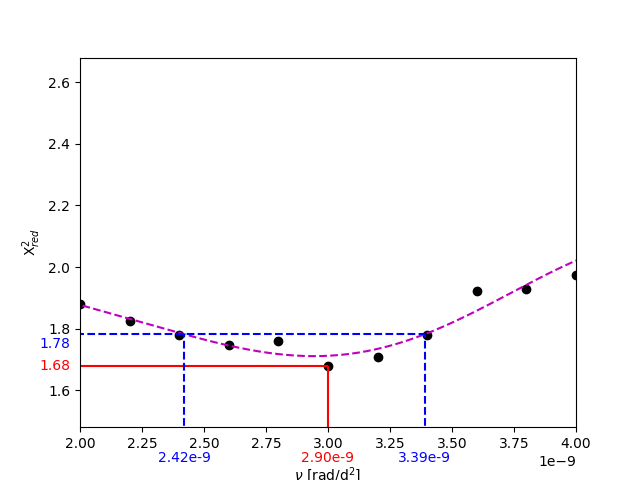}
    \caption{Variation of $\chi_\mathrm{red}^{2}$ of the fit for different models of (2100) Ra-Shalom, keeping fixed the best pole solution and varying $\upsilon$ from 2 to $4\times10^{-9}$. The lowest $\chi_\mathrm{red}^{2}$ value is at $\upsilon = 0.29 \times10^{-8}$, with $\chi_\mathrm{red}^{2}=1.68$ (red solid lines). The $3\sigma$ value  corresponds to $\chi_\mathrm{red}^{2}=1.78$ and is reached at $\upsilon = 0.24 \times10^{-8}$ rad d$^{-2}$ and $\upsilon = 0.34 \times10^{-8}$ rad d$^{-2}$ (blue dashed lines).}
    \label{fig:Ra-Shalom Yorp range}
\end{figure}

 The photometric data set is large ($\sim$ 5000 measurements), so as explained before, we estimated the mean final values of the rotation state parameters ($P$, $\lambda$, $\beta$) with their uncertainties repeating the modelling around the best solution with 100 subsets, removing 25\% of the points in each subset. For the constant period model of Ra-Shalom model we found $P$ $19.820056 \pm 0.000012$ hr, $\lambda = 278^{\circ} \pm 18^{\circ}$, $\beta = -60^{\circ} \pm 8^{\circ}$, $\epsilon = 162^{\circ} \pm 10^{\circ}$ and for the linear increasing period model we found $P=19.820107 \pm 0.000040$ hr, $\lambda = 278^{\circ} \pm 8^{\circ}$, $\beta = -60^{\circ} \pm 5^{\circ}$, $\epsilon = 165^{\circ} \pm 5^{\circ}$ and a YORP acceleration of $\upsilon = (0.22\pm0.16)\times10^{-8}$ rad d$^{-2}$.
We also estimated the uncertainty of the YORP effect in the event it is present at the $3\sigma$ level iterating the YORP code with all parameters, besides the YORP effect, fixed in the previous best solution. In this particular case, we decided to run it from 0 to $0.5\times10^{-8}$ in $0.02\times10^{-8}$ steps, in accordance with the low $\upsilon$ value  derived from the computed model. With this method, we obtain $\upsilon =(0.29\pm0.05)\times10^{-8}$ rad d$^{-2}$ (see Figure \ref{fig:Ra-Shalom Yorp range}).
.

Following \cite{2013MNRAS.430.1376R}, it is possible to estimate the expected YORP acceleration acting on a NEA from a statistical approach knowing its diameter (in km), semi-major axis (in AU) and eccentricity computing $|d\omega/dt| = 1.20^{+1.66}_{-0.86} \times 10^{-2} (a^2\sqrt{1 - e^2}D^2)^{-1}$. Adopting for Ra-Shalom a mean diameter of D=1.76 km from NEOWISE data \citep{2021PSJ.....2..162M}, a semi-major axis $a=0.8321 AU$ and eccentricity $e=0.4365$, we obtain an estimated value for the YORP acceleration of $\nu=4.7^{+6.5}_{-3.3} \times 10^{-8}$ rad d$^{-2}$, one order of magnitude greater that the estimated value from the linearly increasing period code. If we use the diameter estimated from radar physical models \citep{2008Icar..193...20S} of D=2.9 km, we obtain an estimate of $\nu=1.7^{+2.4}_{-1.2} \times 10^{-8}$ rad d$^{-2}$, which is again one order of magnitude greater than our obtained value. Obviously, more observations are necessary to confirm or discard our preliminary result.
Anyway, for our estimated value of $\nu$ it is worth computing the characteristic timescale $T_\mathrm{yorp}$ $=\omega / \nu$, which is the time needed to change the rotation rate of the asteroid significantly. We find that Ra-Shalom may decrease its rotation period to one-half ($\sim$ 10 hr) in about 400 Myr. As this rotation rate is well above the breakup limit, (2100) Ra-Shalom should not experience structural changes in the next 500 Myr due to this effect.

Both linear increasing period and constant period models are a good fit with the data, being slightly better considering an acceleration of the period. It is believed that the YORP effect is responsible of the bimodality in the rotation periods observed in small asteroids, showing greater populations of fast and slow rotators \citep{2000Icar..148...12P}. Interestingly, all asteroids with reported YORP effect to date show acceleration, which could be a bias since they all have fast rotation periods and are therefore easier to study. However, Ra-Shalom is a case of interest because it has a considerably slower rotation period ($\sim$19 hours). Yet, the data suggests an acceleration instead of deceleration, being deceleration a result that would not be unusual given its slow rotation rate. This could also suggest that the YORP effect is more efficient at accelerating than decelerating \citep{2013Icar..225..141S}.
Another hint of the presence of this effect on Ra-Shalom is the value of the ecliptic latitude for its spin pole; we know that another consequence of this effect is to bring the rotation axis to extreme obliquity values \citep{2013A&A...551A..67H}, so a value of $\epsilon = 165^{\circ}$ suggests that this effect could be taking place.

\subsection{(3103) Eger}
\label{subsec:Eger}
Previous studies have focused on (3103) Eger \citep{2009DPS....41.5604D,2012A&A...547A..10D,2018A&A...609A..86D}, detecting the presence of the YORP effect. The most recent study \citep{2018A&A...609A..86D} reports the following rotation state parameters: $P=5.710156 \pm 0.000007$ hr, $\lambda = 226^{\circ}\pm15^{\circ}$, $\beta = -70^{\circ}\pm4^{\circ}$ and $\upsilon = (1.4\pm0.6)\times10^{-8}$ rad d$^{-2}$, from a total of 72 dense lightcurves. In this work we used our ten new lightcurves (see Table \ref{tab_obs_cir}) along with 80 archival lightcurves published by \cite{1987Icar...70..566W,1991Icar...90..117W}, \cite{1992ATsir1553...37V}, \cite{1998Icar..136..124P}, \cite{2012A&A...547A..10D,2018A&A...609A..86D} and \cite{2017MPBu...44..223W} (see Table \ref{tab:Eger archive} for a summary of the archival lightcurves). 

\begin{figure}
    \centering
    \includegraphics[width=\linewidth,keepaspectratio]{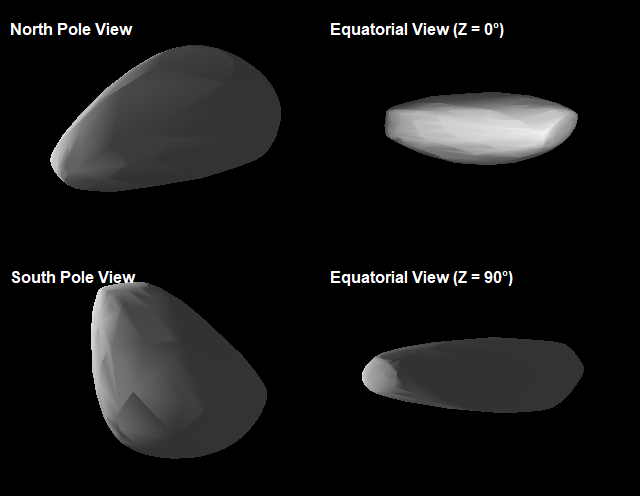}
    \caption{Linearly increasing rotation period shape model of (3103) Eger. Left top: North Pole View (Y axis = 0$^{\circ}$). Left bottom: South Pole View (Y axis = 180$^{\circ}$). Right top and bottom: Equatorial Views with Z axis rotated 0$^{\circ}$ and 90$^{\circ}$.}
    \label{fig:Eger Yorp model}
\end{figure}

\begin{figure*}
    \centering
    \begin{subfigure}{0.33\textwidth}
        \centering
        \includegraphics[width=\textwidth]{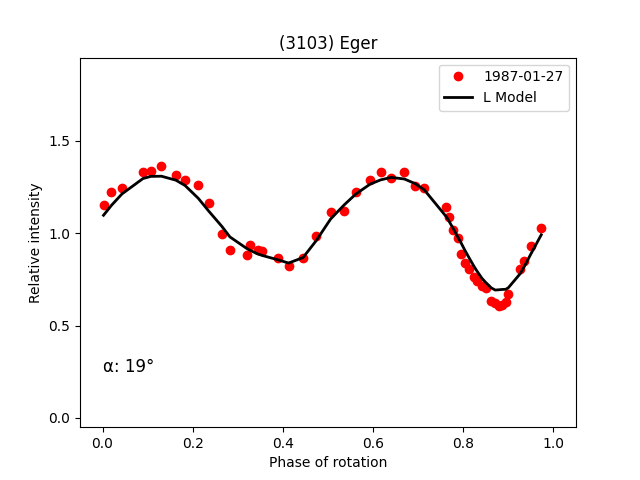}
    \end{subfigure}
    \begin{subfigure}{0.33\textwidth}
        \centering
        \includegraphics[width=\textwidth]{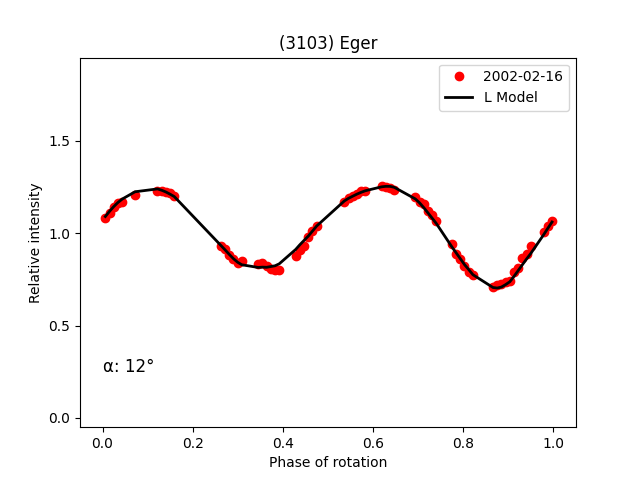}
    \end{subfigure}
    \begin{subfigure}{0.33\textwidth}
        \centering
        \includegraphics[width=\textwidth]{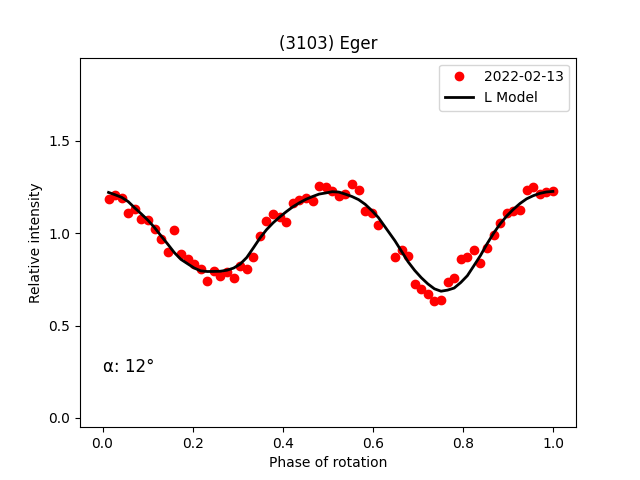}
    \end{subfigure}
    \begin{subfigure}{0.33\textwidth}
        \centering
        \includegraphics[width=\textwidth]{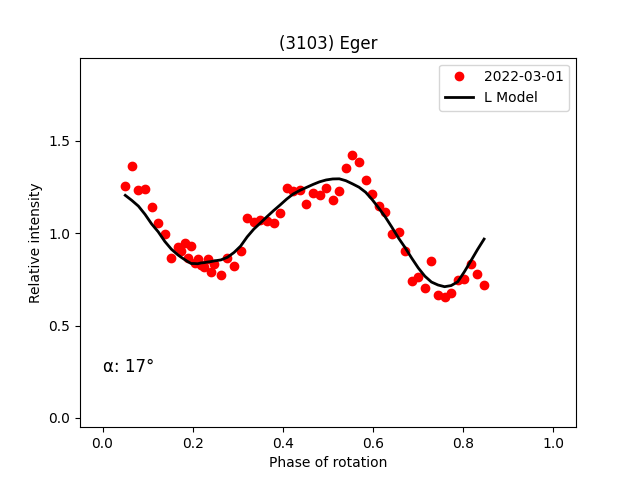}
    \end{subfigure}
    \caption{Four examples of the fit between dense lightcurves of (3103) Eger and the best-fitting linearly increasing period model (L Model). The data is plotted as red dots for each observation, meanwhile the model is plotted as a solid black line. The geometry is described by its solar phase angle $\alpha$.}
    \label{fig:Fit Eger}
\end{figure*}

We computed a model with the YORP code since the effect was already reported. For that, we used 90 lightcurves with a temporal span of 36 years (1986 - 2022), finding as best solution: $\lambda = 214^{\circ}$, $\beta = -71^{\circ}$, $\epsilon \simeq 177^{\circ}$, rotation period corresponding to July 6 1986 (date of the very first observation in the data set) $P = 5.710148$ hr, and a YORP acceleration $\upsilon = 0.847\times10^{-8}$ rad d$^{-2}$. The fit between model and data corresponds to a value of $\chi_\mathrm{red}^{2}=1.74$ normalized to the 6034 data points (see Figures \ref{fig:Fit Eger} and \ref{fig:Eger pole plot}). In Figure \ref{fig:Eger Yorp model} we show the shape model of (3103) Eger.

We recomputed the model around the best solution with 100 sub-sets, each removing 25\% of the points ($\sim$ 6000 measurements). We obtained the following final values: $P=5.710148 \pm 0.000006$ hr, $\lambda = 214^{\circ}\pm 3^{\circ}$, $\beta = -71^{\circ} \pm 1^{\circ}$, $\epsilon = 177^{\circ}\pm 1^{\circ}$ and YORP acceleration $\upsilon = (0.85\pm0.05)\times10^{-8}$ rad d$^{-2}$.

\begin{figure}
    \centering
    \includegraphics[width=\linewidth,keepaspectratio]{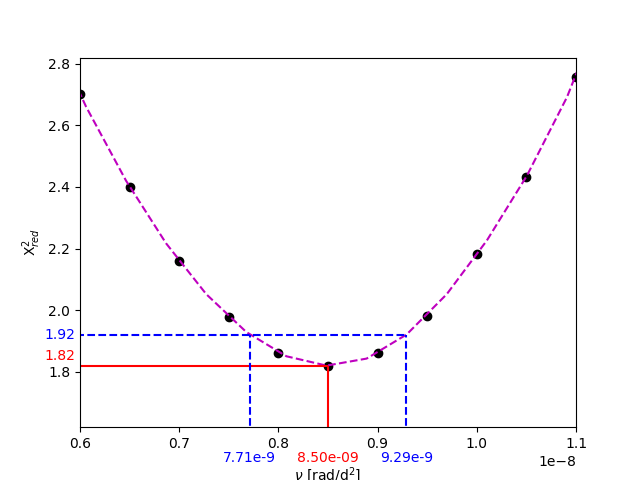}
    \caption{Variation of $\chi_\mathrm{red}^{2}$  of the fit for different models of (3103) Eger, keeping fixed the best pole solution and varying $\upsilon$ from 0.6 to $1.1\times10^{-8}$. The lowest $\chi_\mathrm{red}^{2}$ value is at $\upsilon = 0.85 \times10^{-8}$, with $\chi_\mathrm{red}^{2}=1.82$ (red solid lines). The $3\sigma$ value  corresponds to $\chi_\mathrm{red}^{2}=1.92$ which is reached at $\upsilon = 0.77 \times10^{-8}$ rad d$^{-2}$ and $\upsilon = 0.93 \times10^{-8}$ rad d$^{-2}$(blue dashed lines).}
    \label{fig:Eger Yorp range}
\end{figure}

We employed also the $3\sigma$ method to obtain a second estimation of the uncertainty of $\upsilon$, iterating the $\upsilon$ value  from 0 to $3\times10^{-8}$ in $0.05\times10^{-8}$ steps, and maintaining the rest of the values fixed at the best solution values (see Figure \ref{fig:Eger Yorp range}). In this way we obtained $\upsilon = (0.85\pm0.08)\times10^{-8}$ rad d$^{-2}$, which is in agreement with the previous computed value.

\begin{figure*}
    \centering
    \begin{subfigure}{0.49\linewidth}
        \centering
        \includegraphics[width=\linewidth,keepaspectratio]{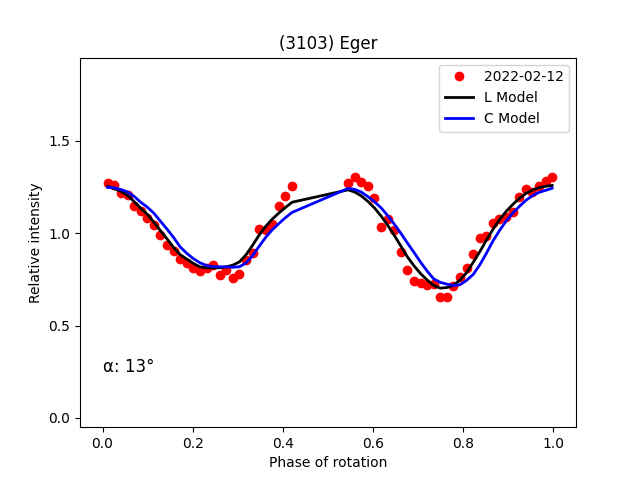}
    \end{subfigure}
    \begin{subfigure}{0.49\linewidth}
        \centering
        \includegraphics[width=\linewidth,keepaspectratio]{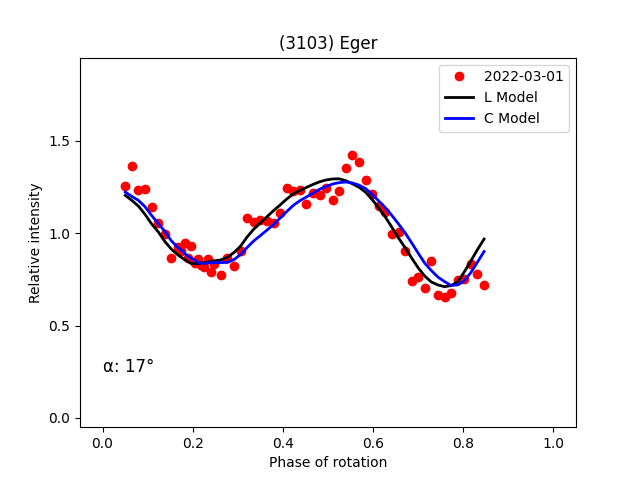}
    \end{subfigure}
    \caption{Example of lightcurves showing the offset of the fit of constant period model (C Model) to both the linearly increasing period model (L model) and the data for (3103) Eger. The data is plotted as red dots for each observation, meanwhile the C Model is plotted as a solid black line and the L Model as a solid blue line. The geometry is described by its solar phase angle $\alpha$}
    \label{fig:Eger lightcurve Yorp and No Yorp}
\end{figure*}

We also computed a shape model with constant period obtaining the following values: $\lambda = 218^{\circ}$, $\beta = -71^{\circ}$, $\epsilon \simeq 178^{\circ}$, rotation period $P=5.710136$ hr with $\chi_\mathrm{red}^{2}=2.95$ (Figure \ref{fig:Eger lightcurve Yorp and No Yorp} shows the fit of both models to some example lightcurves). The $\chi_\mathrm{red}^{2}$ value is higher than the linearly increasing period shape model solution ($\chi_\mathrm{red}^{2}=1.74$) previously obtained, thus we conclude that our linearly increasing period model for (3103) Eger confirms and refines the previous values for its spin parameters and their uncertainties.

For (3103) Eger we estimated a value $T_\mathrm{yorp} = \omega / \nu$ of  $\sim 8$ Myr, time it would take the asteroid to decrease its rotation period to  $\sim$ 2.8 hr, close to the critical rotation period of $\sim$ 2 hr, meaning that significant structural changes could take place within this typical time scale.

\subsection{(12711) Tukmit}
\label{subsec:Tukmit}

Previous studies of this NEA only measured its rotation period, obtaining $P=3.4848 \pm 0.0001$ hr in \citet{2022MPBu...49...83W} and Pravec (2000web)\footnote{https://www.asu.cas.cz/~ppravec/newres.txt}. With our three new dense lightcurves (see Table \ref{tab_obs_cir}), and two archival lighcurves from ALCDEF (see Table \ref{tab:Tukmit archive}), we derived the first spin and shape model for (12711) Tukmit.

\begin{figure}
    \centering
    \includegraphics[width=\linewidth,keepaspectratio]{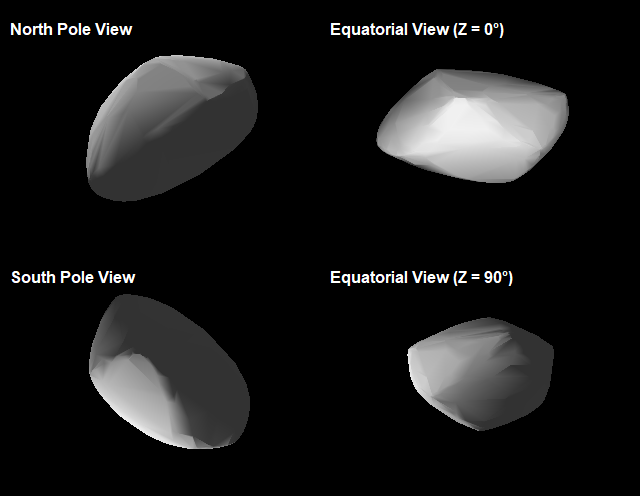}
    \caption{Constant rotation period shape model of (12711) Tukmit. Left top: North Pole View (Y axis = 0$^{\circ}$). Left bottom: South Pole View (Y axis = 180$^{\circ}$). Right top and bottom: Equatorial Views with Z axis rotated 0$^{\circ}$ and 90$^{\circ}$.}
    \label{fig:Tukmit model}
\end{figure}

\begin{figure*}
    \centering
    \begin{subfigure}{0.33\textwidth}
        \centering
        \includegraphics[width=\textwidth]{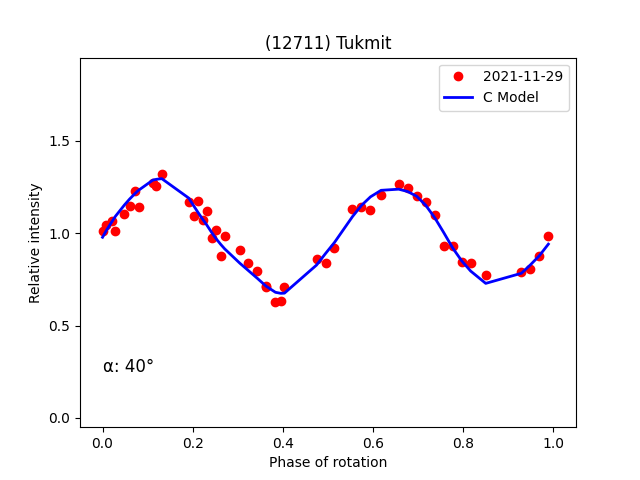}
    \end{subfigure}
    \begin{subfigure}{0.33\textwidth}
        \centering
        \includegraphics[width=\textwidth]{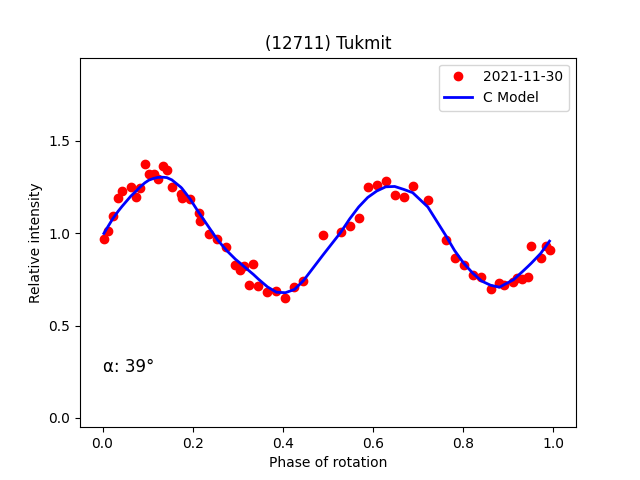}
    \end{subfigure}
    \begin{subfigure}{0.33\textwidth}
        \centering
        \includegraphics[width=\textwidth]{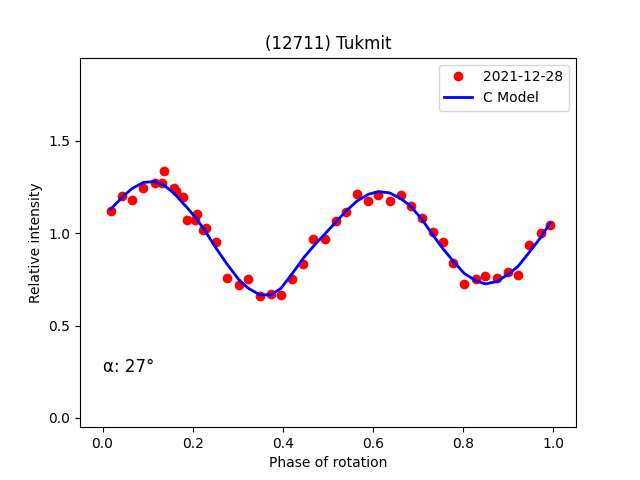}
    \end{subfigure}
    \begin{subfigure}{0.33\textwidth}
        \centering
        \includegraphics[width=\textwidth]{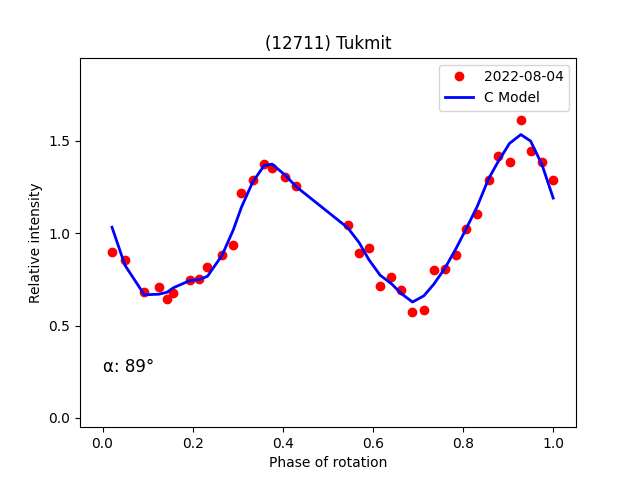}
    \end{subfigure}
    \begin{subfigure}{0.33\textwidth}
        \centering
        \includegraphics[width=\textwidth]{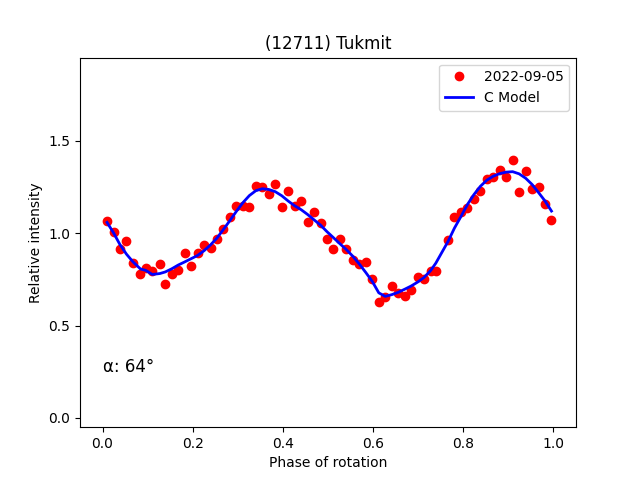}
    \end{subfigure}
    \caption{Fits between all the lightcurves of (12711) Tukmit with the best-fitting constant period model (C Model). The data is plotted as red dots for each observation, meanwhile the model is plotted as a solid blue line. The geometry is described by its solar phase angle $\alpha$.}
    \label{fig:Fit Tukmit}
\end{figure*}

 Due to the short temporal window of the observations (less than one year), we computed a constant period model, obtaining a period of $P=3.484895$ hr with a pole orientation $\lambda = 27^{\circ}$, $\beta = 11^{\circ}$ and $\epsilon \simeq 118^{\circ}$. In Figure \ref{fig:Tukmit model} we show the shape model for this solution. The fit between model and data has in this case $\chi_\mathrm{red}^{2}=1.06$ (see Figures \ref{fig:Fit Tukmit} and \ref{fig:Tukmit pole plot}).

To estimate the mean values and their uncertainties, since the main data set for Tukmit is smaller compared to the others ($\sim$ 150 measurements), we decided to remove 10\% of the main data to obtain each subset instead of 25\%. We obtained $P=3.484900 \pm 0.000031$ hr $\lambda = 27^{\circ}\pm 8^{\circ}$, $\beta = 9^{\circ} \pm 15^{\circ}$ and $\epsilon = 119^{\circ}\pm 15^{\circ}$.

Since the time span of the observations is so small ($\sim$ 1 year) it is extremely unlikely that we would detect the YORP effect, if it were present, unless being extremely strong. Anyway, we computed a linear increasing period model, but as expected, the obtained best-fitting model was unsuccessful to improve the constant $P$ model. We note that the aforementioned obliquity expected in a YORP affected asteroid is not present in the best-fitting model obtained ($\epsilon \simeq 118^{\circ}$)).
Anyway, according to \cite{2013MNRAS.430.1376R} we could expect a YORP acceleration of $\nu=1.8^{+2.5}_{-1.3} \times 10^{-8}$ rad d$^{-2}$, assuming D=1.94 km \citep{2010AJ....140..770T}, a=1.1863 AU and e= 0.2721. If so, the value $T_\mathrm{yorp} = \omega / \nu$ would be $\sim 8$ Myr, time at which the asteroid would reach a rotation period of $\sim$ 1.7 hr, well beyond the critical rotation limit. More observations are needed to confirm and refine our results for (12711) Tukmit.

\subsection{(161989) Cacus}
\label{subsec:Cacus}

This asteroid has been already studied in \cite{2018A&A...609A..86D}, being reported to be affected by YORP. The published parameters are $P=3.755067 \pm 0.000002$ hr (for the first observation of February 28 1978), $\lambda = 254^{\circ}\pm5^{\circ}$, $\beta = -62^{\circ}\pm2^{\circ}$ and $\upsilon = (1.9\pm0.3)\times10^{-8}$ rad d$^{-2}$. To compute that model a set of 22 lightcurves was used (see Table \ref{tab:Cacus archive}), spanning from 1978 to 2016.

\begin{figure}
    \centering
    \includegraphics[width=\linewidth,keepaspectratio]{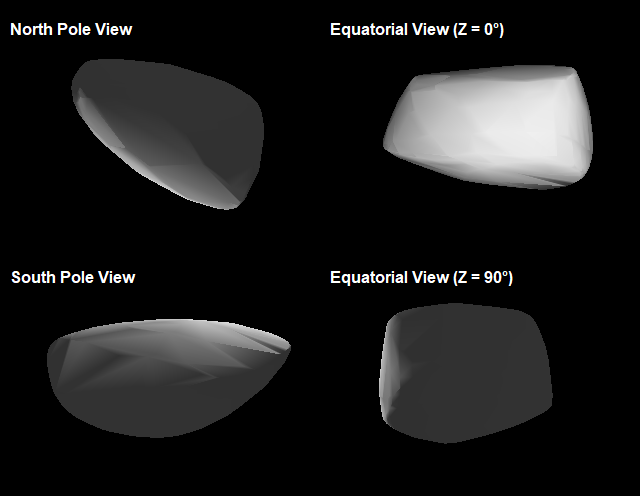}
    \caption{Linearly increasing rotation period shape model of (161989) Cacus. Left top: North Pole View (Y axis = 0$^{\circ}$). Left bottom: South Pole View (Y axis = 180$^{\circ}$). Right top and bottom: Equatorial Views with Z axis rotated 0$^{\circ}$ and 90$^{\circ}$.}
    \label{fig:Cacus Yorp model}
\end{figure}

\begin{figure*}
    \centering
    \begin{subfigure}{0.33\textwidth}
        \centering
        \includegraphics[width=\textwidth]{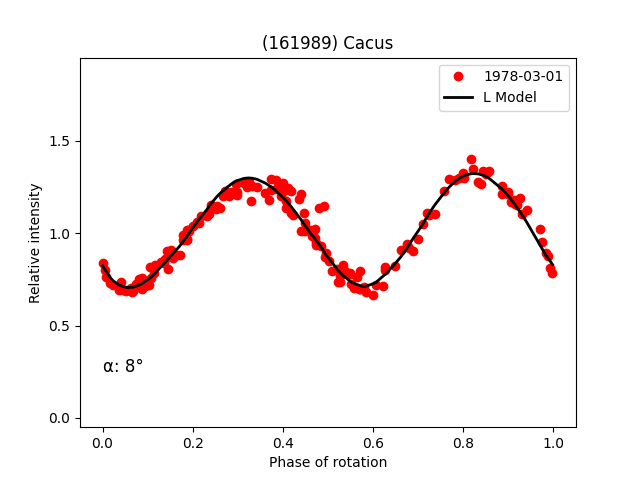}
    \end{subfigure}
    \begin{subfigure}{0.33\textwidth}
        \centering
        \includegraphics[width=\textwidth]{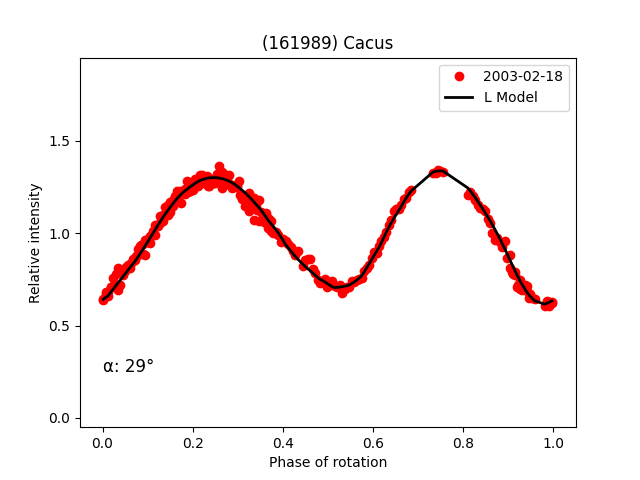}
    \end{subfigure}
    \begin{subfigure}{0.33\textwidth}
        \centering
        \includegraphics[width=\textwidth]{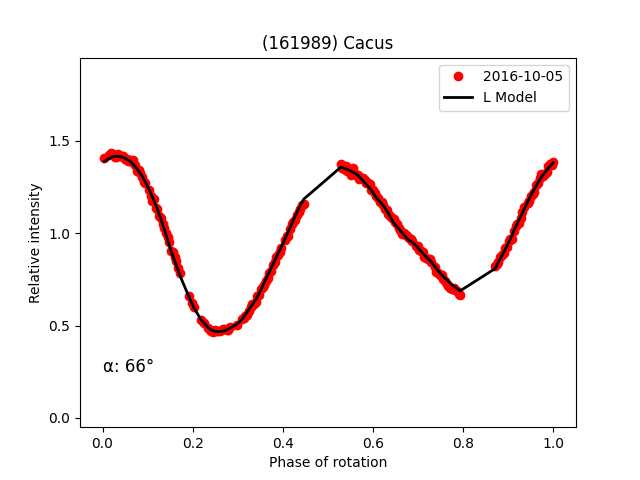}
    \end{subfigure}
    \begin{subfigure}{0.33\textwidth}
        \centering
        \includegraphics[width=\textwidth]{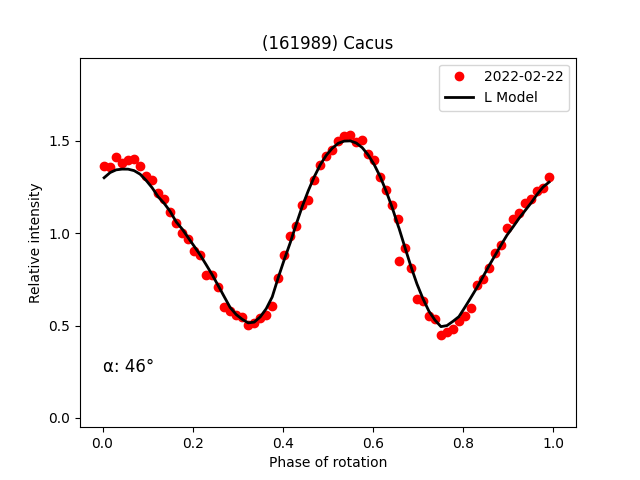}
    \end{subfigure}
    \begin{subfigure}{0.33\textwidth}
        \centering
        \includegraphics[width=\textwidth]{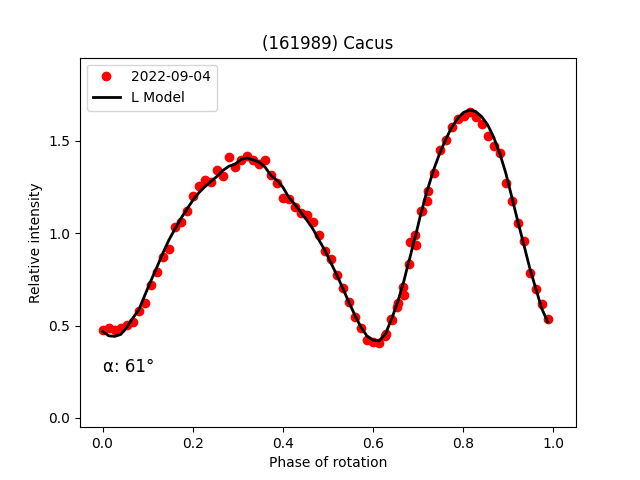}
    \end{subfigure}
    \caption{Fits between five lightcurves of (161989) Cacus and the best-fitting linearly increasing period model (L Model). The data is plotted as red dots for each observation, meanwhile the model is plotted as a solid black line. The geometry is described by its solar phase angle $\alpha$.}
    \label{fig:Fit Cacus}
\end{figure*}

We added to those previous observations our three new lightcurves acquired during 2022 (see Table \ref{tab_obs_cir}), increasing to 44 years the temporal window of the observations. We computed a linearly increasing period model since the YORP effect has been previously reported for (161989) Cacus. The best-fitting solution  corresponds to a pole orientation of  $\lambda = 251^{\circ}$, $\beta = -61^{\circ}$, $\epsilon \simeq 178^{\circ}$, $P=3.755067$ hr (corresponding to February 28 1978) and a YORP acceleration $\upsilon = 1.91\times10^{-8}$ rad d$^{-2}$. The fit between the model and data corresponds to a value of $\chi_\mathrm{red}^{2}=1.31$ normalized to the 1534 data points. In Figure \ref{fig:Cacus Yorp model} we show the associated shape model (see Figure \ref{fig:Fit Cacus} for a graphical representation of the fit).

To obtain the final mean values and their uncertainties for each parameter of the model, we recomputed the model for 100 subsets obtained removing randomly 25\% of the data from the main set (in this case the number of measurements is large enough $\sim$ 1500 measurements). We obtained $P=3.755067 \pm 0.000001$ hr, $\lambda = 251^{\circ}\pm 6^{\circ}$, $\beta = -62^{\circ} \pm 2^{\circ}$, $\epsilon = 177^{\circ}\pm 2^{\circ}$  and $\upsilon = (1.91\pm0.05)\times10^{-8}$ rad d$^{-2}$.

\begin{figure}
    \centering
    \includegraphics[width=\linewidth,keepaspectratio]{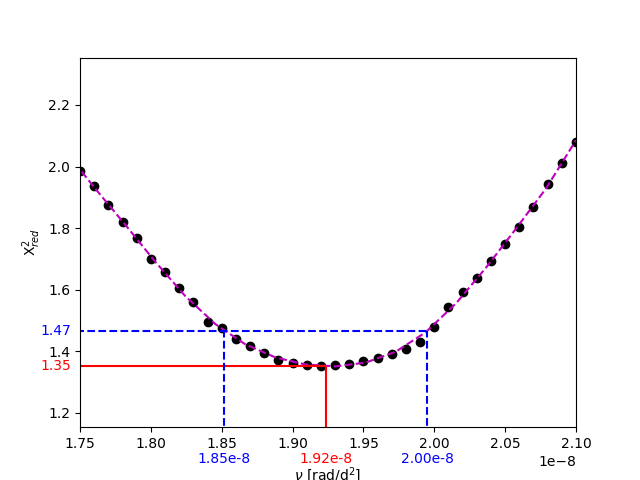}
    \caption{Variation of $\chi_\mathrm{red}^{2}$ of the fit for different models of (161989) Cacus, keeping fixed the best pole solution and varying $\upsilon$ from 1.75 to $2.1\times10^{-8}$. The lowest $\chi_\mathrm{red}^{2}$ value is at $\upsilon = 1.92 \times10^{-8}$, with $\chi_\mathrm{red}^{2}=1.35$ (red solid lines). The $3\sigma$ value  corresponds to $\chi_\mathrm{red}^{2}=1.47$ which is reached at $\upsilon = 1.85 \times10^{-8}$ rad d$^{-2}$ and $\upsilon = 2.00 \times10^{-8}$ rad d$^{-2}$(blue dashed lines).}
    \label{fig:Cacus Yorp range}
\end{figure}

We also used the $3\sigma$ method to estimate the uncertainty of the YORP effect, iterating in this case the $\upsilon$ value between 0 and $3\times10^{-8}$ with $0.01\times10^{-8}$ steps. In this way we find $\upsilon = (1.92\pm0.08)\times10^{-8}$ rad d$^{-2}$ (see Figures \ref{fig:Cacus Yorp range} and \ref{fig:Cacus pole plot}), in good agreement with the best-fitting model.

\begin{figure*}
    \centering
    \begin{subfigure}{0.49\linewidth}
        \centering
        \includegraphics[width=\linewidth,keepaspectratio]{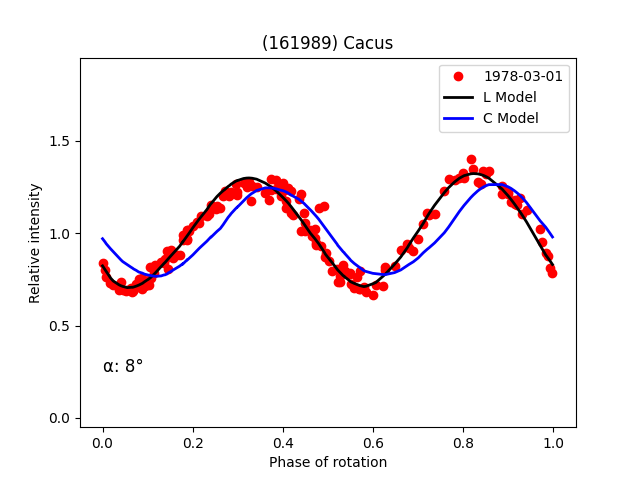}
    \end{subfigure}
    \begin{subfigure}{0.49\linewidth}
        \centering
        \includegraphics[width=\linewidth,keepaspectratio]{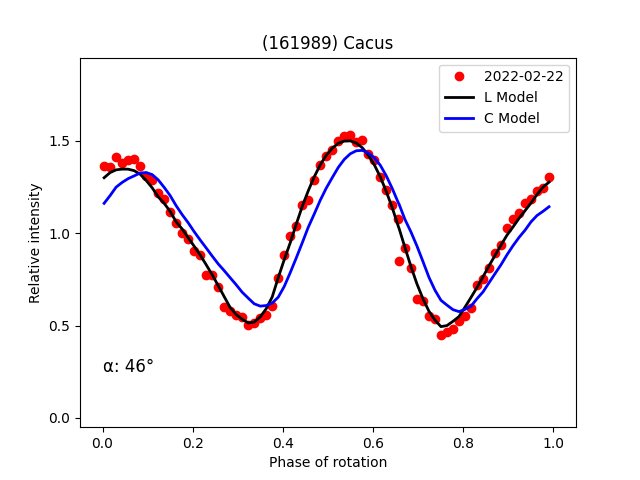}
    \end{subfigure}
    \caption{Example of lightcurves showing the offset of the fit of constant period model (C Model) to both the linearly increasing period model (L model) and the data for (161989) Cacus. The data is plotted as red dots for each observation, meanwhile the C Model is plotted as a solid black line and the L Model as a solid blue line. The geometry is described by its solar phase angle $\alpha$}
    \label{fig:Cacus lightcurve Yorp and No Yorp}
\end{figure*}

As for (3103) Eger, we also computed a shape model with constant period, obtaining the following values: $\lambda = 245^{\circ}$, $\beta = -61^{\circ}$, $\epsilon \simeq 176^{\circ}$ rotation period $P=3.755052$ hr and $\chi_\mathrm{red}^{2}=13.65$ (Figure \ref{fig:Cacus lightcurve Yorp and No Yorp} shows the fit of both models to some example lightcurves). The $\chi_\mathrm{red}^{2}$ value is much higher than the linearly increasing period shape model ($\chi_\mathrm{red}^{2}=1.31$) previously obtained, thus, we conclude that our results for (161989) Cacus confirm previous works and significantly decrease the uncertainty of the $\upsilon$ value.

We also estimate $T_\mathrm{yorp} = \omega / \nu$  $\sim 8.2$ Myr, time scale at which the asteroid would reach a rotation period of $\sim 1.9$ hr that is beyond the critical rotation period.  

\section{Conclusions}
\label{sec:Conclussions}
In this work, we computed models, spin state and shape, including period changes due to YORP for asteroids (2100) Ra-Shalom, (3103) Eger, (12711) Tukmit and (161989) Cacus. For asteroids (3103) Eger and (161989) Cacus, our results agree with those published by \cite{2012A&A...547A..10D,2018A&A...609A..86D}, obtaining smaller uncertainties. For (3103) Eger we found $P=5.710148 \pm 0.000006$ hr, $\lambda = 214^{\circ}\pm 3^{\circ}$, $\beta = -71^{\circ} \pm 1^{\circ}$, $\epsilon = 177^{\circ}\pm 1^{\circ}$ and YORP acceleration $\upsilon = (0.85\pm0.05)\times10^{-8}$ rad d$^{-2}$. For (161989) Cacus our best-fitting rotation state parameters are: $P=3.755067 \pm 0.000001$ hr, $\lambda = 251^{\circ}\pm 6^{\circ}$, $\beta = -62^{\circ} \pm 2^{\circ}$, $\epsilon = 177^{\circ}\pm 2^{\circ}$ and a YORP acceleration $\upsilon = (1.91\pm0.05)\times10^{-8}$ rad d$^{-2}$.

For (2100) Ra-Shalom, while the rotation state parameters ($P$, $\lambda$, $\beta$) agree with the results proposed in \cite{2018A&A...609A..86D}, we can not discard a hint of YORP acceleration taking place, since the best-fitting model with linearly increasing rotation period has a slightly lower $\chi_\mathrm{red}^{2}$ value and uncertainties than the constant period model. We obtained using a constant period model: $P = 19.820056 \pm 0.000012$ hr, $\lambda = 278^{\circ} \pm 18^{\circ}$, $\beta = -60^{\circ} \pm 8^{\circ}$ and $\epsilon = 162^{\circ} \pm 10^{\circ}$, meanwhile the values obtained for this asteroid with a linear increasing period are: $\lambda = 278^{\circ} \pm 8^{\circ}$, $\beta = -60^{\circ} \pm 5^{\circ}$, $\epsilon = 165^{\circ} \pm 5^{\circ}$ with a rotation period of $P=19.820107 \pm 0.000040$ hr and YORP acceleration $\upsilon = (0.22\pm0.16)\times10^{-8}$ rad d$^{-2}$. It is also worth mentioning that to compute the uncertainties a 100 models were created in a $30^{\circ}$ x $30^{\circ}$ square centered around the best-fitting solution, obtaining values near the solution and always positive. If so, (161989) Ra-Shalom would be the slowest rotator of the known asteroids with YORP detection. Furthermore, this could also be a hint that this effect is more effective accelerating than decelerating.

Finally, for asteroid (12711) Tukmit we present the first shape model and rotation state parameters ($P$, $\lambda$, $\beta$) from a limited set of lightcurves, confirming and refining the period published by \cite{2022MPBu...49...83W}, and finding $P=3.484900 \pm 0.000031$ hr, $ \lambda = 27^{\circ}\pm 8^{\circ}$, $\beta = 9^{\circ} \pm 15^{\circ}$ and $\epsilon = 119^{\circ}\pm 15^{\circ}$.

\section*{Acknowledgements}

We thank Dr. Josef {\v{D}}urech for providing us the inversion code that includes the YORP acceleration and for his advises in using the inversion codes. This project has received funding from the European Union's Horizon 2020 research and innovation program under grant agreement No 870403 (NEOROCKS). JL, MRA and MS-R acknowledge support from the ACIISI, Consejer\'{\i}a de Econom\'{\i}a, Conocimiento y Empleo del Gobierno de Canarias and the European Regional Development Fund (ERDF) under grant with reference ProID2021010134 and support from the Agencia Estatal de Investigacion del Ministerio de Ciencia e Innovacion (AEI-MCINN) under grant "Hydrated Minerals and Organic Compounds in Primitive Asteroids" with reference PID2020-120464GB-100. 
This reseach was also funded by FICYT (FUNDACION PARA LA INVESTIGACION CIENTIFICA Y TECNICA), grant number SV-PA-21-AYUD/2021/51301 and Plan Nacional by Ministerio de Ciencia, Innovación y Universidades, Spain, grant number MCIU-22-PID2021-127331NB-I00

This article is based on observations made with the Telescopio IAC80 and TAR2 telescopes operated on the island of Tenerife by the Instituto de Astrof\'{\i}sica de Canarias in the Spanish Observatorio del Teide.

The work has been funded by HUNOSA through the collaboration agreement with reference SV-21-HUNOSA-2.

This work uses data obtained from the Asteroid Lightcurve Data Exchange Format (ALCDEF) database, which is supported by funding from NASA grant 80NSSC18K0851. 
\section*{Data Availability}

 The data underlying this article will be shared on reasonable request to the corresponding author.




\bibliographystyle{mnras}
\bibliography{bibliography} 

\begin{thebibliography}{}
\makeatletter
\relax
\def\mn@urlcharsother{\let\do\@makeother \do\$\do\&\do\#\do\^\do\_\do\%\do\~}
\def\mn@doi{\begingroup\mn@urlcharsother \@ifnextchar [ {\mn@doi@} {\mn@doi@[]}}
\def\mn@doi@[#1]#2{\def\@tempa{#1}\ifx\@tempa\@empty \href {http://dx.doi.org/#2} {doi:#2}\else \href {http://dx.doi.org/#2} {#1}\fi \endgroup}
\def\mn@eprint#1#2{\mn@eprint@#1:#2::\@nil}
\def\mn@eprint@arXiv#1{\href {http://arxiv.org/abs/#1} {{\tt arXiv:#1}}}
\def\mn@eprint@dblp#1{\href {http://dblp.uni-trier.de/rec/bibtex/#1.xml} {dblp:#1}}
\def\mn@eprint@#1:#2:#3:#4\@nil{\def\@tempa {#1}\def\@tempb {#2}\def\@tempc {#3}\ifx \@tempc \@empty \let \@tempc \@tempb \let \@tempb \@tempa \fi \ifx \@tempb \@empty \def\@tempb {arXiv}\fi \@ifundefined {mn@eprint@\@tempb}{\@tempb:\@tempc}{\expandafter \expandafter \csname mn@eprint@\@tempb\endcsname \expandafter{\@tempc}}}

\bibitem[\protect\citeauthoryear{{Alarcon}, {Licandro}, {Serra-Ricart}, {Joven}, {Gaitan}  \& {de Sousa}}{{Alarcon} et~al.}{2023}]{2023arXiv230203700A}
{Alarcon} M.~R.,  {Licandro} J.,  {Serra-Ricart} M.,  {Joven} E.,  {Gaitan} V.,   {de Sousa} R.,  2023, \mn@doi [arXiv e-prints] {10.48550/arXiv.2302.03700}, \href {https://ui.adsabs.harvard.edu/abs/2023arXiv230203700A} {p. arXiv:2302.03700}

\bibitem[\protect\citeauthoryear{{Benner} et~al.,}{{Benner} et~al.}{1997}]{1997Icar..130..296B}
{Benner} L. A.~M.,  et~al., 1997, \mn@doi [\icarus] {10.1006/icar.1997.5834}, \href {https://ui.adsabs.harvard.edu/abs/1997Icar..130..296B} {130, 296}

\bibitem[\protect\citeauthoryear{{Benner} et~al.,}{{Benner} et~al.}{2008}]{2008Icar..198..294B}
{Benner} L. A.~M.,  et~al., 2008, \mn@doi [\icarus] {10.1016/j.icarus.2008.06.010}, \href {https://ui.adsabs.harvard.edu/abs/2008Icar..198..294B} {198, 294}

\bibitem[\protect\citeauthoryear{{Bottke}, {Vokrouhlick{\'y}}, {Rubincam}  \& {Nesvorn{\'y}}}{{Bottke} et~al.}{2006}]{2006AREPS..34..157B}
{Bottke} William~F. J.,  {Vokrouhlick{\'y}} D.,  {Rubincam} D.~P.,   {Nesvorn{\'y}} D.,  2006, \mn@doi [Annual Review of Earth and Planetary Sciences] {10.1146/annurev.earth.34.031405.125154}, \href {https://ui.adsabs.harvard.edu/abs/2006AREPS..34..157B} {34, 157}

\bibitem[\protect\citeauthoryear{{Chesley} et~al.,}{{Chesley} et~al.}{2003}]{2003Sci...302.1739C}
{Chesley} S.~R.,  et~al., 2003, \mn@doi [Science] {10.1126/science.1091452}, \href {https://ui.adsabs.harvard.edu/abs/2003Sci...302.1739C} {302, 1739}

\bibitem[\protect\citeauthoryear{{Degewij}, {Lebofsky}  \& {Lebofsky}}{{Degewij} et~al.}{1978}]{1978IAUC.3193....1D}
{Degewij} J.,  {Lebofsky} L.,   {Lebofsky} M.,  1978, \iaucirc, \href {https://ui.adsabs.harvard.edu/abs/1978IAUC.3193....1D} {3193, 1}

\bibitem[\protect\citeauthoryear{{Hanu{\v{s}}} et~al.,}{{Hanu{\v{s}}} et~al.}{2011}]{2011A&A...530A.134H}
{Hanu{\v{s}}} J.,  et~al., 2011, \mn@doi [\aap] {10.1051/0004-6361/201116738}, \href {https://ui.adsabs.harvard.edu/abs/2011A&A...530A.134H} {530, A134}

\bibitem[\protect\citeauthoryear{{Hanu{\v{s}}} et~al.,}{{Hanu{\v{s}}} et~al.}{2013a}]{2013A&A...551A..67H}
{Hanu{\v{s}}} J.,  et~al., 2013a, \mn@doi [\aap] {10.1051/0004-6361/201220701}, \href {https://ui.adsabs.harvard.edu/abs/2013A&A...551A..67H} {551, A67}

\bibitem[\protect\citeauthoryear{{Hanu{\v{s}}} et~al.,}{{Hanu{\v{s}}} et~al.}{2013b}]{2013A&A...559A.134H}
{Hanu{\v{s}}} J.,  et~al., 2013b, \mn@doi [\aap] {10.1051/0004-6361/201321993}, \href {https://ui.adsabs.harvard.edu/abs/2013A&A...559A.134H} {559, A134}

\bibitem[\protect\citeauthoryear{{Harris}, {Young}, {Dockweiler}, {Gibson}, {Poutanen}  \& {Bowell}}{{Harris} et~al.}{1992}]{1992Icar...95..115H}
{Harris} A.~W.,  {Young} J.~W.,  {Dockweiler} T.,  {Gibson} J.,  {Poutanen} M.,   {Bowell} E.,  1992, \mn@doi [\icarus] {10.1016/0019-1035(92)90195-D}, \href {https://ui.adsabs.harvard.edu/abs/1992Icar...95..115H} {95, 115}

\bibitem[\protect\citeauthoryear{{Heinze} et~al.,}{{Heinze} et~al.}{2018}]{2018AJ....156..241H}
{Heinze} A.~N.,  et~al., 2018, \mn@doi [\aj] {10.3847/1538-3881/aae47f}, \href {https://ui.adsabs.harvard.edu/abs/2018AJ....156..241H} {156, 241}

\bibitem[\protect\citeauthoryear{{Kaasalainen} \& {Torppa}}{{Kaasalainen} \& {Torppa}}{2001}]{2001Icar..153...24K}
{Kaasalainen} M.,  {Torppa} J.,  2001, \mn@doi [\icarus] {10.1006/icar.2001.6673}, \href {https://ui.adsabs.harvard.edu/abs/2001Icar..153...24K} {153, 24}

\bibitem[\protect\citeauthoryear{{Kaasalainen}, {Torppa}  \& {Muinonen}}{{Kaasalainen} et~al.}{2001}]{2001Icar..153...37K}
{Kaasalainen} M.,  {Torppa} J.,   {Muinonen} K.,  2001, \mn@doi [\icarus] {10.1006/icar.2001.6674}, \href {https://ui.adsabs.harvard.edu/abs/2001Icar..153...37K} {153, 37}

\bibitem[\protect\citeauthoryear{{Kaasalainen} et~al.,}{{Kaasalainen} et~al.}{2004}]{2004Icar..167..178K}
{Kaasalainen} M.,  et~al., 2004, \mn@doi [\icarus] {10.1016/j.icarus.2003.09.012}, \href {https://ui.adsabs.harvard.edu/abs/2004Icar..167..178K} {167, 178}

\bibitem[\protect\citeauthoryear{{Kaasalainen}, {{\v{D}}urech}, {Warner}, {Krugly}  \& {Gaftonyuk}}{{Kaasalainen} et~al.}{2007}]{2007Natur.446..420K}
{Kaasalainen} M.,  {{\v{D}}urech} J.,  {Warner} B.~D.,  {Krugly} Y.~N.,   {Gaftonyuk} N.~M.,  2007, \mn@doi [\nat] {10.1038/nature05614}, \href {https://ui.adsabs.harvard.edu/abs/2007Natur.446..420K} {446, 420}

\bibitem[\protect\citeauthoryear{{Kochanek} et~al.,}{{Kochanek} et~al.}{2017}]{2017PASP..129j4502K}
{Kochanek} C.~S.,  et~al., 2017, \mn@doi [\pasp] {10.1088/1538-3873/aa80d9}, \href {https://ui.adsabs.harvard.edu/abs/2017PASP..129j4502K} {129, 104502}

\bibitem[\protect\citeauthoryear{{Koehn}, {Bowell}, {Skiff}, {Sanborn}, {McLelland}, {Pravec}  \& {Warner}}{{Koehn} et~al.}{2014}]{2014MPBu...41..286K}
{Koehn} B.~W.,  {Bowell} E.~G.,  {Skiff} B.~A.,  {Sanborn} J.~J.,  {McLelland} K.~P.,  {Pravec} P.,   {Warner} B.~D.,  2014, Minor Planet Bulletin, \href {https://ui.adsabs.harvard.edu/abs/2014MPBu...41..286K} {41, 286}

\bibitem[\protect\citeauthoryear{{Licandro} et~al.,}{{Licandro} et~al.}{2023}]{2023MNRAS.tmp..693L}
{Licandro} J.,  et~al., 2023, \mn@doi [\mnras] {10.1093/mnras/stad708}, \href {https://ui.adsabs.harvard.edu/abs/2023MNRAS.tmp..693L} {}

\bibitem[\protect\citeauthoryear{{Lowry} et~al.,}{{Lowry} et~al.}{2007}]{2007Sci...316..272L}
{Lowry} S.~C.,  et~al., 2007, \mn@doi [Science] {10.1126/science.1139040}, \href {https://ui.adsabs.harvard.edu/abs/2007Sci...316..272L} {316, 272}

\bibitem[\protect\citeauthoryear{{Lowry} et~al.,}{{Lowry} et~al.}{2014}]{2014A&A...562A..48L}
{Lowry} S.~C.,  et~al., 2014, \mn@doi [\aap] {10.1051/0004-6361/201322602}, \href {https://ui.adsabs.harvard.edu/abs/2014A&A...562A..48L} {562, A48}

\bibitem[\protect\citeauthoryear{{Masiero}, {Mainzer}, {Bauer}, {Cutri}, {Grav}, {Kramer}, {Pittichov{\'a}}  \& {Wright}}{{Masiero} et~al.}{2021}]{2021PSJ.....2..162M}
{Masiero} J.~R.,  {Mainzer} A.~K.,  {Bauer} J.~M.,  {Cutri} R.~M.,  {Grav} T.,  {Kramer} E.,  {Pittichov{\'a}} J.,   {Wright} E.~L.,  2021, \mn@doi [\psj] {10.3847/PSJ/ac15fb}, \href {https://ui.adsabs.harvard.edu/abs/2021PSJ.....2..162M} {2, 162}

\bibitem[\protect\citeauthoryear{{Mommert}}{{Mommert}}{2017}]{2017A&C....18...47M}
{Mommert} M.,  2017, \mn@doi [Astronomy and Computing] {10.1016/j.ascom.2016.11.002}, \href {https://ui.adsabs.harvard.edu/abs/2017A&C....18...47M} {18, 47}

\bibitem[\protect\citeauthoryear{{Morbidelli} \& {Vokrouhlick{\'y}}}{{Morbidelli} \& {Vokrouhlick{\'y}}}{2003}]{2003Icar..163..120M}
{Morbidelli} A.,  {Vokrouhlick{\'y}} D.,  2003, \mn@doi [\icarus] {10.1016/S0019-1035(03)00047-2}, \href {https://ui.adsabs.harvard.edu/abs/2003Icar..163..120M} {163, 120}

\bibitem[\protect\citeauthoryear{O'Keefe}{O'Keefe}{1976}]{o1976tektites}
O'Keefe J.,  1976, Tektites and Their Origin, 161

\bibitem[\protect\citeauthoryear{{Ostro}, {Harris}, {Campbell}, {Shapiro}  \& {Young}}{{Ostro} et~al.}{1984}]{1984Icar...60..391O}
{Ostro} S.~J.,  {Harris} A.~W.,  {Campbell} D.~B.,  {Shapiro} I.~I.,   {Young} J.~W.,  1984, \mn@doi [\icarus] {10.1016/0019-1035(84)90198-2}, \href {https://ui.adsabs.harvard.edu/abs/1984Icar...60..391O} {60, 391}

\bibitem[\protect\citeauthoryear{{Paddack}}{{Paddack}}{1969}]{1969JGR....74.4379P}
{Paddack} S.~J.,  1969, \mn@doi [\jgr] {10.1029/JB074i017p04379}, \href {https://ui.adsabs.harvard.edu/abs/1969JGR....74.4379P} {74, 4379}

\bibitem[\protect\citeauthoryear{{Parley} et~al.,}{{Parley} et~al.}{2005}]{2005EM&P...97..261P}
{Parley} N.~R.,  et~al., 2005, \mn@doi [Earth Moon and Planets] {10.1007/s11038-006-9072-z}, \href {https://ui.adsabs.harvard.edu/abs/2005EM&P...97..261P} {97, 261}

\bibitem[\protect\citeauthoryear{{Pravec} \& {Harris}}{{Pravec} \& {Harris}}{2000}]{2000Icar..148...12P}
{Pravec} P.,  {Harris} A.~W.,  2000, \mn@doi [\icarus] {10.1006/icar.2000.6482}, \href {https://ui.adsabs.harvard.edu/abs/2000Icar..148...12P} {148, 12}

\bibitem[\protect\citeauthoryear{{Pravec}, {Wolf}  \& {{\v{S}}arounov{\'a}}}{{Pravec} et~al.}{1998}]{1998Icar..136..124P}
{Pravec} P.,  {Wolf} M.,   {{\v{S}}arounov{\'a}} L.,  1998, \mn@doi [\icarus] {10.1006/icar.1998.5993}, \href {https://ui.adsabs.harvard.edu/abs/1998Icar..136..124P} {136, 124}

\bibitem[\protect\citeauthoryear{{Pravec} et~al.,}{{Pravec} et~al.}{2008}]{2008Icar..197..497P}
{Pravec} P.,  et~al., 2008, \mn@doi [\icarus] {10.1016/j.icarus.2008.05.012}, \href {https://ui.adsabs.harvard.edu/abs/2008Icar..197..497P} {197, 497}

\bibitem[\protect\citeauthoryear{{Radzievskii}}{{Radzievskii}}{1952}]{1952AZh....29..162R}
{Radzievskii} V.~V.,  1952, \azh, \href {https://ui.adsabs.harvard.edu/abs/1952AZh....29..162R} {29, 162}

\bibitem[\protect\citeauthoryear{{Rozitis} \& {Green}}{{Rozitis} \& {Green}}{2013}]{2013MNRAS.430.1376R}
{Rozitis} B.,  {Green} S.~F.,  2013, \mn@doi [\mnras] {10.1093/mnras/sts723}, \href {https://ui.adsabs.harvard.edu/abs/2013MNRAS.430.1376R} {430, 1376}

\bibitem[\protect\citeauthoryear{{Schuster}, {Surdej}  \& {Surdej}}{{Schuster} et~al.}{1979}]{1979A&AS...37..483S}
{Schuster} H.~E.,  {Surdej} A.,   {Surdej} J.,  1979, \aaps, \href {https://ui.adsabs.harvard.edu/abs/1979A&AS...37..483S} {37, 483}

\bibitem[\protect\citeauthoryear{{Shepard}, {Benner}, {Ostro}, {Harris}, {Rosema}, {Shapiro}, {Chandler}  \& {Campbell}}{{Shepard} et~al.}{2000}]{2000Icar..147..520S}
{Shepard} M.~K.,  {Benner} L. A.~M.,  {Ostro} S.~J.,  {Harris} A.~W.,  {Rosema} K.~D.,  {Shapiro} I.~I.,  {Chandler} J.~F.,   {Campbell} D.~B.,  2000, \mn@doi [\icarus] {10.1006/icar.2000.6470}, \href {https://ui.adsabs.harvard.edu/abs/2000Icar..147..520S} {147, 520}

\bibitem[\protect\citeauthoryear{{Shepard} et~al.,}{{Shepard} et~al.}{2008a}]{2008Icar..193...20S}
{Shepard} M.~K.,  et~al., 2008a, \mn@doi [\icarus] {10.1016/j.icarus.2007.09.006}, \href {https://ui.adsabs.harvard.edu/abs/2008Icar..193...20S} {193, 20}

\bibitem[\protect\citeauthoryear{{Shepard} et~al.,}{{Shepard} et~al.}{2008b}]{2008Icar..195..184S}
{Shepard} M.~K.,  et~al., 2008b, \mn@doi [\icarus] {10.1016/j.icarus.2007.11.032}, \href {https://ui.adsabs.harvard.edu/abs/2008Icar..195..184S} {195, 184}

\bibitem[\protect\citeauthoryear{{Slivan}}{{Slivan}}{2002}]{2002Natur.419...49S}
{Slivan} S.~M.,  2002, \mn@doi [\nat] {10.1038/nature00993}, \href {https://ui.adsabs.harvard.edu/abs/2002Natur.419...49S} {419, 49}

\bibitem[\protect\citeauthoryear{{Statler}, {Cotto-Figueroa}, {Riethmiller}  \& {Sweeney}}{{Statler} et~al.}{2013}]{2013Icar..225..141S}
{Statler} T.~S.,  {Cotto-Figueroa} D.,  {Riethmiller} D.~A.,   {Sweeney} K.~M.,  2013, \mn@doi [\icarus] {10.1016/j.icarus.2013.03.010}, \href {https://ui.adsabs.harvard.edu/abs/2013Icar..225..141S} {225, 141}

\bibitem[\protect\citeauthoryear{{Stephens} \& {Warner}}{{Stephens} \& {Warner}}{2018}]{2018DPS....5041703S}
{Stephens} R.,  {Warner} B.~D.,  2018, in AAS/Division for Planetary Sciences Meeting Abstracts \#50. p. 417.03

\bibitem[\protect\citeauthoryear{{Taylor} et~al.,}{{Taylor} et~al.}{2007}]{2007Sci...316..274T}
{Taylor} P.~A.,  et~al., 2007, \mn@doi [Science] {10.1126/science.1139038}, \href {https://ui.adsabs.harvard.edu/abs/2007Sci...316..274T} {316, 274}

\bibitem[\protect\citeauthoryear{{Tonry} et~al.,}{{Tonry} et~al.}{2018}]{2018PASP..130f4505T}
{Tonry} J.~L.,  et~al., 2018, \mn@doi [\pasp] {10.1088/1538-3873/aabadf}, \href {https://ui.adsabs.harvard.edu/abs/2018PASP..130f4505T} {130, 064505}

\bibitem[\protect\citeauthoryear{{Torppa}, {Kaasalainen}, {Micha{\l}owski}, {Kwiatkowski}, {Kryszczy{\'n}ska}, {Denchev}  \& {Kowalski}}{{Torppa} et~al.}{2003}]{2003Icar..164..346T}
{Torppa} J.,  {Kaasalainen} M.,  {Micha{\l}owski} T.,  {Kwiatkowski} T.,  {Kryszczy{\'n}ska} A.,  {Denchev} P.,   {Kowalski} R.,  2003, \mn@doi [\icarus] {10.1016/S0019-1035(03)00146-5}, \href {https://ui.adsabs.harvard.edu/abs/2003Icar..164..346T} {164, 346}

\bibitem[\protect\citeauthoryear{{Trilling} et~al.,}{{Trilling} et~al.}{2010}]{2010AJ....140..770T}
{Trilling} D.~E.,  et~al., 2010, \mn@doi [\aj] {10.1088/0004-6256/140/3/770}, \href {https://ui.adsabs.harvard.edu/abs/2010AJ....140..770T} {140, 770}

\bibitem[\protect\citeauthoryear{{Velichko}, {Kruglyj}  \& {Chernyj}}{{Velichko} et~al.}{1992}]{1992ATsir1553...37V}
{Velichko} F.~P.,  {Kruglyj} Y.~N.,   {Chernyj} V.~G.,  1992, Astronomicheskij Tsirkulyar, \href {https://ui.adsabs.harvard.edu/abs/1992ATsir1553...37V} {1553, 37}

\bibitem[\protect\citeauthoryear{{Vokrouhlick{\'y}}, {Bottke}, {Chesley}, {Scheeres}  \& {Statler}}{{Vokrouhlick{\'y}} et~al.}{2015}]{2015aste.book..509V}
{Vokrouhlick{\'y}} D.,  {Bottke} W.~F.,  {Chesley} S.~R.,  {Scheeres} D.~J.,   {Statler} T.~S.,  2015, in , Asteroids IV.
pp 509--531, \mn@doi{10.2458/azu_uapress_9780816532131-ch027}

\bibitem[\protect\citeauthoryear{{Vokrouhlick{\'y}} et~al.,}{{Vokrouhlick{\'y}} et~al.}{2017}]{2017AJ....153..270V}
{Vokrouhlick{\'y}} D.,  et~al., 2017, \mn@doi [\aj] {10.3847/1538-3881/aa72ea}, \href {https://ui.adsabs.harvard.edu/abs/2017AJ....153..270V} {153, 270}

\bibitem[\protect\citeauthoryear{{Walsh}, {Richardson}  \& {Michel}}{{Walsh} et~al.}{2008}]{2008Natur.454..188W}
{Walsh} K.~J.,  {Richardson} D.~C.,   {Michel} P.,  2008, \mn@doi [\nat] {10.1038/nature07078}, \href {https://ui.adsabs.harvard.edu/abs/2008Natur.454..188W} {454, 188}

\bibitem[\protect\citeauthoryear{{Warner}}{{Warner}}{2017}]{2017MPBu...44..223W}
{Warner} B.~D.,  2017, Minor Planet Bulletin, \href {https://ui.adsabs.harvard.edu/abs/2017MPBu...44..223W} {44, 223}

\bibitem[\protect\citeauthoryear{{Warner} \& {Stephens}}{{Warner} \& {Stephens}}{2022}]{2022MPBu...49...83W}
{Warner} B.~D.,  {Stephens} R.~D.,  2022, Minor Planet Bulletin, \href {https://ui.adsabs.harvard.edu/abs/2022MPBu...49...83W} {49, 83}

\bibitem[\protect\citeauthoryear{{Wisniewski}}{{Wisniewski}}{1987}]{1987Icar...70..566W}
{Wisniewski} W.~Z.,  1987, \mn@doi [\icarus] {10.1016/0019-1035(87)90096-0}, \href {https://ui.adsabs.harvard.edu/abs/1987Icar...70..566W} {70, 566}

\bibitem[\protect\citeauthoryear{{Wisniewski}}{{Wisniewski}}{1991}]{1991Icar...90..117W}
{Wisniewski} W.~Z.,  1991, \mn@doi [\icarus] {10.1016/0019-1035(91)90073-3}, \href {https://ui.adsabs.harvard.edu/abs/1991Icar...90..117W} {90, 117}

\bibitem[\protect\citeauthoryear{Yarkovsky}{Yarkovsky}{1901}]{yarkovsky1901density}
Yarkovsky I.,  1901, Bryansk, published privately by the author

\bibitem[\protect\citeauthoryear{{{\v{D}}urech} et~al.,}{{{\v{D}}urech} et~al.}{2007}]{2007A&A...465..331D}
{{\v{D}}urech} J.,  et~al., 2007, \mn@doi [\aap] {10.1051/0004-6361:20066347}, \href {https://ui.adsabs.harvard.edu/abs/2007A&A...465..331D} {465, 331}

\bibitem[\protect\citeauthoryear{{{\v{D}}urech} et~al.,}{{{\v{D}}urech} et~al.}{2008}]{2008A&A...489L..25D}
{{\v{D}}urech} J.,  et~al., 2008, \mn@doi [\aap] {10.1051/0004-6361:200810672}, \href {https://ui.adsabs.harvard.edu/abs/2008A&A...489L..25D} {489, L25}

\bibitem[\protect\citeauthoryear{{{\v{D}}urech} et~al.,}{{{\v{D}}urech} et~al.}{2009a}]{2009DPS....41.5604D}
{{\v{D}}urech} J.,  et~al., 2009a, in AAS/Division for Planetary Sciences Meeting Abstracts \#41. p. 56.04

\bibitem[\protect\citeauthoryear{{{\v{D}}urech} et~al.,}{{{\v{D}}urech} et~al.}{2009b}]{2009A&A...493..291D}
{{\v{D}}urech} J.,  et~al., 2009b, \mn@doi [\aap] {10.1051/0004-6361:200810393}, \href {https://ui.adsabs.harvard.edu/abs/2009A&A...493..291D} {493, 291}

\bibitem[\protect\citeauthoryear{{{\v{D}}urech}, {Sidorin}  \& {Kaasalainen}}{{{\v{D}}urech} et~al.}{2010}]{2010A&A...513A..46D}
{{\v{D}}urech} J.,  {Sidorin} V.,   {Kaasalainen} M.,  2010, \mn@doi [\aap] {10.1051/0004-6361/200912693}, \href {https://ui.adsabs.harvard.edu/abs/2010A&A...513A..46D} {513, A46}

\bibitem[\protect\citeauthoryear{{{\v{D}}urech} et~al.,}{{{\v{D}}urech} et~al.}{2012}]{2012A&A...547A..10D}
{{\v{D}}urech} J.,  et~al., 2012, \mn@doi [\aap] {10.1051/0004-6361/201219396}, \href {https://ui.adsabs.harvard.edu/abs/2012A&A...547A..10D} {547, A10}

\bibitem[\protect\citeauthoryear{{{\v{D}}urech}, {Hanu{\v{s}}}, {Oszkiewicz}  \& {Van{\v{c}}o}}{{{\v{D}}urech} et~al.}{2016}]{2016A&A...587A..48D}
{{\v{D}}urech} J.,  {Hanu{\v{s}}} J.,  {Oszkiewicz} D.,   {Van{\v{c}}o} R.,  2016, \mn@doi [\aap] {10.1051/0004-6361/201527573}, \href {https://ui.adsabs.harvard.edu/abs/2016A&A...587A..48D} {587, A48}

\bibitem[\protect\citeauthoryear{{{\v{D}}urech} et~al.,}{{{\v{D}}urech} et~al.}{2018}]{2018A&A...609A..86D}
{{\v{D}}urech} J.,  et~al., 2018, \mn@doi [\aap] {10.1051/0004-6361/201731465}, \href {https://ui.adsabs.harvard.edu/abs/2018A&A...609A..86D} {609, A86}

\bibitem[\protect\citeauthoryear{{{\v{D}}urech}, {Hanu{\v{s}}}  \& {Van{\v{c}}o}}{{{\v{D}}urech} et~al.}{2019}]{2019A&A...631A...2D}
{{\v{D}}urech} J.,  {Hanu{\v{s}}} J.,   {Van{\v{c}}o} R.,  2019, \mn@doi [\aap] {10.1051/0004-6361/201936341}, \href {https://ui.adsabs.harvard.edu/abs/2019A&A...631A...2D} {631, A2}

\makeatother
\end{thebibliography}




\appendix

\section{Summary of archival lightcurves used in this work}

\begin{table*}
\caption{Archival observations for (2100) Ra-Shalom. The information includes the date, the starting and end time (UT) of the observations, the phase angle ($\alpha$), the heliocentric ($r$) and geocentric ($\Delta$) distances, phase angle bisector longitude (PABLon) and latitude (PABLat) of the asteroid at the time of observation. \textbf{References:} HAR92: \protect\cite{1992Icar...95..115H}; OST84: \protect\cite{1984Icar...60..391O}; PRA98: \protect\cite{1998Icar..136..124P}; KAA04: \protect\cite{2004Icar..167..178K}; DUR12: \protect\cite{2012A&A...547A..10D}; DUR18: \protect\cite{2018A&A...609A..86D}.}
\label{tab:Ra-Shalom archive}
\begin{tabular}{lllrrrrrl}
\hline
 Date        & UT (start)   & UT (end)    &   $\alpha [^\circ]$ &   $r$ [au] &   $\Delta$ [au] &   PABLon [deg] &   PABLat [deg] & Reference   \\
 \hline
 
 1978-Sep-12 & 05:35:59.971 & 0:12:00.000 &                3.06 &     1.1945 &          0.1884 &       348.549  &         1.9168 & HAR92           \\
 1981-Aug-25 & 05:42:09.504 & 1:36:28.224 &               30.63 &     1.1624 &          0.1811 &       331.617  &        20.5476 & OST84           \\
 1981-Aug-28 & 08:43:20.352 & 0:58:24.672 &               28.07 &     1.1696 &          0.1851 &       329.838  &        18.231  & OST84           \\
 1981-Sep-02 & 03:47:02.688 & 8:58:09.408 &               27.63 &     1.1789 &          0.1965 &       327.606  &        14.644  & OST84           \\
 1997-Aug-30 & 21:34:59.002 & 3:02:03.638 &               41.27 &     1.1952 &          0.2677 &         8.1616 &         2.7398 & PRA98           \\
 1997-Sep-01 & 21:58:07.190 & 2:55:50.995 &               39.06 &     1.1949 &          0.256  &         8.1911 &         1.8904 & PRA98           \\
 1997-Sep-02 & 21:35:20.602 & 3:09:22.550 &               37.93 &     1.1945 &          0.2504 &         8.1712 &         1.4562 & PRA98           \\
 1997-Sep-03 & 23:02:59.251 & 3:14:06.979 &               36.65 &     1.1941 &          0.2445 &         8.1231 &         0.9737 & PRA98           \\
 1997-Sep-06 & 00:28:43.853 & 3:22:17.126 &               34.02 &     1.193  &          0.2333 &         7.9446 &        -0.0099 & PRA98           \\
 1997-Sep-11 & 21:23:44.304 & 3:35:41.165 &               25.57 &     1.1876 &          0.2051 &         6.7514 &        -3.1789 & PRA98           \\
 2000-Aug-23 & 19:55:14.246 & 3:23:03.754 &               34.82 &     1.1791 &          0.2137 &       351.287  &        13.6909 & KAA04           \\
 2000-Aug-24 & 00:26:36.672 & 5:56:32.986 &               34.49 &     1.1794 &          0.213  &       351.217  &        13.6175 & KAA04           \\
 2000-Aug-24 & 22:46:46.301 & 4:26:49.142 &               32.81 &     1.1809 &          0.21   &       350.867  &        13.233  & KAA04           \\
 2000-Aug-25 & 20:57:49.795 & 2:45:19.325 &               31.1  &     1.1824 &          0.2072 &       350.494  &        12.8392 & KAA04           \\
 2000-Aug-26 & 19:35:11.818 & 2:52:43.939 &               29.32 &     1.1838 &          0.2045 &       350.088  &        12.4257 & KAA04           \\
 2000-Aug-27 & 04:34:47.280 & 5:53:59.280 &               28.6  &     1.1843 &          0.2035 &       349.913  &        12.2629 & KAA04           \\
 2003-Aug-06 & 19:17:10.003 & 0:46:24.499 &               63.99 &     1.0826 &          0.1881 &       333.888  &        37.2542 & DUR12           \\
 2003-Aug-24 & 00:03:29.952 & 0:22:04.166 &               37.87 &     1.1474 &          0.1802 &       323.434  &        24.2987 & DUR12           \\
 2003-Aug-24 & 21:57:08.179 & 1:25:21.101 &               37.15 &     1.15   &          0.182  &       323.002  &        23.4851 & DUR12           \\
 2003-Aug-25 & 21:39:35.136 & 0:50:24.778 &               36.51 &     1.1528 &          0.1842 &       322.561  &        22.6059 & DUR12           \\
 2003-Aug-27 & 17:52:42.902 & 0:16:37.229 &               35.69 &     1.1578 &          0.189  &       321.831  &        20.9753 & DUR12           \\
 2003-Aug-29 & 17:46:55.661 & 0:51:28.800 &               35.32 &     1.1628 &          0.1952 &       321.173  &        19.2443 & DUR12           \\
 2003-Aug-30 & 18:03:19.930 & 0:04:23.952 &               35.32 &     1.1653 &          0.1988 &       320.894  &        18.3859 & DUR12           \\
 2003-Aug-31 & 23:51:33.264 & 0:18:39.312 &               35.49 &     1.1681 &          0.2034 &       320.593  &        17.3552 & DUR12           \\
 2003-Sep-02 & 21:41:22.272 & 0:08:19.997 &               36    &     1.1722 &          0.2113 &       320.245  &        15.8157 & DUR12           \\
 2003-Sep-05 & 21:15:57.312 & 3:08:13.747 &               37.31 &     1.1779 &          0.225  &       319.928  &        13.5496 & DUR12           \\
 2003-Sep-06 & 22:08:17.952 & 3:12:01.930 &               37.87 &     1.1797 &          0.2302 &       319.878  &        12.8038 & DUR12           \\
 2003-Sep-14 & 20:06:33.955 & 1:01:11.885 &               42.75 &     1.1901 &          0.2747 &       320.347  &         7.7938 & DUR12           \\
 2003-Sep-15 & 18:26:53.088 & 0:25:59.837 &               43.33 &     1.191  &          0.2804 &       320.485  &         7.2794 & DUR12           \\
 2003-Sep-16 & 18:21:37.210 & 0:01:37.517 &               43.94 &     1.1918 &          0.2866 &       320.647  &         6.7446 & DUR12           \\
 2003-Sep-17 & 18:51:25.085 & 9:50:50.726 &               44.56 &     1.1925 &          0.293  &       320.829  &         6.2131 & DUR12           \\
 2009-Aug-13 & 17:37:01.315 & 9:54:49.277 &               84.86 &     0.9792 &          0.3617 &       262.483  &        30.6297 & DUR12           \\
 2009-Aug-14 & 17:35:55.046 & 9:44:51.130 &               83.83 &     0.9855 &          0.3625 &       263.831  &        30.0153 & DUR12           \\
 2009-Aug-16 & 17:28:54.538 & 9:32:55.306 &               81.8  &     0.9979 &          0.3651 &       266.449  &        28.7684 & DUR12           \\
 2009-Aug-17 & 17:15:11.491 & 9:57:09.245 &               80.81 &     1.0039 &          0.3668 &       267.712  &        28.1429 & DUR12           \\
 2009-Aug-23 & 18:38:14.957 & 0:30:10.656 &               75.21 &     1.0385 &          0.3831 &       274.927  &        24.3233 & DUR12           \\
 2009-Sep-19 & 16:30:22.723 & 9:14:26.275 &               60.83 &     1.1501 &          0.5405 &       298.924  &        10.782  & DUR12           \\
 2009-Sep-20 & 16:28:47.510 & 9:08:04.474 &               60.55 &     1.1529 &          0.548  &       299.644  &        10.4035 & DUR12           \\
 2009-Sep-21 & 16:31:15.341 & 8:58:07.018 &               60.28 &     1.1556 &          0.5555 &       300.358  &        10.0314 & DUR12           \\
 2013-Sep-07 & 00:01:06.096 & 3:14:18.038 &               59.18 &     1.1529 &          0.4025 &        33.4915 &        -8.2415 & DUR18           \\
 2013-Sep-08 & 00:00:07.603 & 3:08:08.419 &               59.33 &     1.1501 &          0.3952 &        34.1256 &        -8.7121 & DUR18           \\
 2013-Sep-10 & 00:17:10.061 & 1:09:01.411 &               59.68 &     1.1441 &          0.3806 &        35.4226 &        -9.7022 & DUR18           \\
 2013-Sep-27 & 01:35:02.314 & 3:40:39.158 &               66.78 &     1.0783 &          0.2738 &        47.9848 &       -21.1684 & DUR18           \\
 2013-Sep-28 & 01:58:03.677 & 3:54:32.746 &               67.55 &     1.0735 &          0.2689 &        48.8524 &       -22.0693 & DUR18           \\
 2016-Aug-10 & 08:30:02.966 & 1:13:57.619 &               57.23 &     1.1907 &          0.4862 &        10.1379 &         5.1135 & DUR18           \\
 2016-Aug-11 & 08:26:16.512 & 1:41:30.538 &               56.98 &     1.1916 &          0.4794 &        10.5629 &         4.8588 & DUR18           \\
 2016-Aug-12 & 08:20:44.822 & 1:52:10.762 &               56.73 &     1.1923 &          0.4725 &        10.9843 &         4.6001 & DUR18           \\
 2016-Aug-13 & 08:25:49.987 & 1:29:16.483 &               56.46 &     1.193  &          0.4655 &        11.4055 &         4.3347 & DUR18           \\
 2016-Aug-14 & 08:21:19.814 & 1:49:09.667 &               56.2  &     1.1936 &          0.4585 &        11.8208 &         4.0662 & DUR18           \\
 2016-Aug-15 & 08:14:47.299 & 1:59:42.634 &               55.93 &     1.1941 &          0.4515 &        12.2322 &         3.7929 & DUR18           \\
 2016-Aug-16 & 08:46:39.418 & 2:03:16.646 &               55.64 &     1.1945 &          0.4443 &        12.6509 &         3.5067 & DUR18           \\
 2016-Aug-19 & 09:16:34.205 & 0:51:48.701 &               54.76 &     1.1952 &          0.4228 &        13.8668 &         2.6249 & DUR18           \\
 2016-Aug-20 & 09:13:49.786 & 1:36:21.053 &               54.45 &     1.1952 &          0.4157 &        14.2611 &         2.3207 & DUR18           \\
 2016-Aug-25 & 23:03:40.896 & 2:59:53.693 &               52.6  &     1.1938 &          0.3757 &        16.3893 &         0.4809 & DUR18           \\
 2016-Aug-27 & 22:55:21.763 & 3:00:50.112 &               51.86 &     1.1926 &          0.3614 &        17.1111 &        -0.2421 & DUR18           \\
 2016-Aug-29 & 23:49:11.741 & 2:51:28.771 &               51.06 &     1.191  &          0.3467 &        17.8244 &        -1.0227 & DUR18           \\
 2016-Aug-30 & 23:03:34.589 & 2:52:12.922 &               50.67 &     1.1901 &          0.3398 &        18.1539 &        -1.4102 & DUR18           \\
 2016-Sep-02 & 22:46:43.018 & 2:32:33.734 &               49.38 &     1.1868 &          0.3187 &        19.1281 &        -2.6793 & DUR18           \\
 2016-Sep-10 & 00:18:12.010 & 2:31:41.635 &               45.98 &     1.1758 &          0.2702 &        21.1091 &        -6.2176 & DUR18           \\
 2016-Sep-11 & 16:42:55.958 & 8:57:38.678 &               45.1  &     1.1726 &          0.2592 &        21.4923 &        -7.1972 & DUR18           \\
 2016-Sep-16 & 16:17:18.730 & 9:00:05.818 &               42.56 &     1.1613 &          0.2282 &        22.3748 &       -10.4834 & DUR18           \\
 2016-Sep-19 & 17:58:12.518 & 8:50:24.691 &               41.18 &     1.1532 &          0.2106 &        22.6733 &       -12.8409 & DUR18           \\
\hline

\end{tabular}
\end{table*}

\begin{table*}
\caption{Continuation of table \ref{tab:Ra-Shalom archive}}
\begin{tabular}{lllrrrrrl}
\hline
 Date        & UT (start)   & UT (end)    &   $\alpha [^\circ]$ &   $r$ [au] &   $\Delta$ [au] &   PABLon [deg] &   PABLat [deg] & Reference   \\
 \hline
 2016-Sep-22 & 16:36:35.539 & 8:30:08.957 &               40.21 &     1.1447 &          0.1953 &        22.7238 &       -15.3684 & DUR18           \\
 2016-Sep-23 & 16:56:25.526 & 7:53:51.936 &               40    &     1.1415 &          0.1903 &        22.6801 &       -16.3031 & DUR18           \\
 2016-Sep-25 & 17:03:34.502 & 8:49:42.701 &               39.86 &     1.135  &          0.1813 &        22.4863 &       -18.2482 & DUR18           \\
 2016-Sep-26 & 16:43:28.963 & 8:51:45.216 &               39.95 &     1.1317 &          0.1772 &        22.334  &       -19.2504 & DUR18           \\
 2016-Sep-27 & 16:26:26.419 & 8:41:40.934 &               40.15 &     1.1283 &          0.1733 &        22.1416 &       -20.2838 & DUR18           \\
 2016-Oct-08 & 00:16:45.869 & 1:06:03.341 &               50.96 &     1.087  &          0.1503 &        17.4875 &       -32.0972 & DUR18           \\
 2016-Oct-08 & 07:16:18.595 & 8:21:30.787 &               51.46 &     1.0857 &          0.1502 &        17.2791 &       -32.4224 & DUR18           \\
 2016-Oct-08 & 08:17:40.099 & 9:17:18.787 &               51.53 &     1.0855 &          0.1502 &        17.2502 &       -32.4695 & DUR18           \\
 2016-Oct-09 & 01:33:02.650 & 2:31:11.482 &               52.86 &     1.0822 &          0.1499 &        16.7576 &       -33.2909 & DUR18           \\
 2016-Oct-09 & 08:30:40.378 & 9:16:35.674 &               53.38 &     1.0809 &          0.1499 &        16.5441 &       -33.6065 & DUR18           \\
 2016-Oct-10 & 01:12:32.314 & 2:05:34.426 &               54.71 &     1.0777 &          0.1499 &        16.0483 &       -34.3862 & DUR18           \\
 2016-Oct-10 & 02:51:36.634 & 3:45:13.306 &               54.84 &     1.0773 &          0.1499 &        15.9941 &       -34.4603 & DUR18           \\
 2016-Oct-10 & 04:30:26.266 & 5:09:16.474 &               54.97 &     1.077  &          0.15   &        15.9406 &       -34.5333 & DUR18           \\
 2016-Oct-10 & 05:04:28.762 & 5:35:28.954 &               55.02 &     1.0769 &          0.15   &        15.9225 &       -34.5583 & DUR18           \\
 2016-Oct-10 & 07:30:08.986 & 8:03:12.730 &               55.2  &     1.0764 &          0.15   &        15.8469 &       -34.665  & DUR18           \\
 2016-Oct-10 & 08:32:19.738 & 9:13:05.722 &               55.28 &     1.0762 &          0.15   &        15.8157 &       -34.7106 & DUR18           \\
 2016-Oct-13 & 10:17:33.418 & 2:27:43.891 &               61.44 &     1.0613 &          0.1522 &        13.4717 &       -37.9095 & DUR18           \\
 2016-Oct-13 & 12:23:07.411 & 4:23:05.654 &               61.62 &     1.0609 &          0.1524 &        13.409  &       -37.9954 & DUR18           \\
 2016-Oct-13 & 14:16:11.626 & 6:10:06.730 &               61.78 &     1.0605 &          0.1524 &        13.3526 &       -38.0742 & DUR18           \\
 2016-Oct-14 & 11:56:14.582 & 4:06:16.502 &               63.64 &     1.0559 &          0.1537 &        12.6441 &       -38.9464 & DUR18           \\
 2016-Oct-14 & 14:01:40.627 & 6:20:16.541 &               63.82 &     1.0554 &          0.1538 &        12.5818 &       -39.0305 & DUR18           \\
 2016-Oct-14 & 16:21:25.661 & 8:28:47.741 &               64.02 &     1.0549 &          0.1539 &        12.5107 &       -39.1256 & DUR18           \\
 2016-Oct-15 & 09:10:18.365 & 1:14:43.930 &               65.46 &     1.0513 &          0.155  &        11.9518 &       -39.775  & DUR18           \\
 2016-Oct-15 & 11:10:07.709 & 3:16:44.515 &               65.63 &     1.0509 &          0.1552 &        11.8922 &       -39.8503 & DUR18           \\
 2016-Oct-15 & 13:36:17.914 & 5:04:30.950 &               65.84 &     1.0504 &          0.1554 &        11.8204 &       -39.9442 & DUR18           \\
 2016-Oct-17 & 09:29:16.512 & 1:49:29.798 &               69.57 &     1.0407 &          0.1589 &        10.4243 &       -41.553  & DUR18           \\
 2016-Oct-17 & 11:39:08.237 & 2:38:03.034 &               69.75 &     1.0402 &          0.1591 &        10.363  &       -41.6289 & DUR18           \\
 2016-Oct-17 & 15:00:45.446 & 7:55:32.506 &               70.03 &     1.0394 &          0.1595 &        10.2679 &       -41.7501 & DUR18           \\
 2016-Oct-25 & 13:51:48.874 & 7:50:22.243 &               84.98 &     0.9934 &          0.1826 &         5.3987 &       -47.5766 & DUR18           \\
 2016-Oct-26 & 13:37:33.859 & 5:35:19.565 &               86.67 &     0.9872 &          0.1861 &         4.9692 &       -48.2212 & DUR18           \\
 2016-Oct-26 & 15:33:04.262 & 7:39:46.771 &               86.81 &     0.9867 &          0.1864 &         4.9396 &       -48.2762 & DUR18           \\
\hline
\end{tabular}
\end{table*}

\begin{table*}
\caption{Archival observations for (3103) Eger. The information includes the date, the starting and end time (UT) of the observations, the phase angle ($\alpha$), the heliocentric ($r$) and geocentric ($\Delta$) distances, phase angle bisector longitude (PABLon) and latitude (PABLat) of the asteroid at the time of observation. \textbf{References:} WIS87: \protect\cite{1987Icar...70..566W}; VEL92: \protect\cite{1992ATsir1553...37V}; PRA98: \protect\cite{1998Icar..136..124P}; DUR12: \protect\cite{2012A&A...547A..10D}; WAR17: \protect\cite{2017MPBu...44..223W}; DUR18: \protect\cite{2018A&A...609A..86D}.}
\label{tab:Eger archive}

\begin{tabular}{lllrrrrrl}
\hline
 Date        & UT (start)   & UT (end)    &   $\alpha [^\circ]$ &   $r$ [au] &   $\Delta$ [au] &   PABLon [deg] &   PABLat [deg] & Reference   \\
 \hline
 
 1986-Jul-06 & 07:47:25.958 & 0:18:09.706 &               44.24 &     1.2215 &          0.3206 &       316.316  &        14.1376 & WIS87           \\
 1986-Jul-12 & 07:18:13.680 & 0:43:15.485 &               44.95 &     1.1887 &          0.2683 &       321.301  &        11.3073 & WIS87           \\
 1986-Aug-07 & 09:38:39.581 & 2:37:16.032 &               71.09 &     1.0519 &          0.1454 &       352.978  &       -22.2845 & WIS87           \\
 1987-Jan-26 & 06:53:01.248 & 2:54:02.419 &               20.26 &     1.4193 &          0.4783 &       145.334  &         4.1932 & WIS87           \\
 1987-Jan-27 & 06:06:02.016 & 1:54:01.958 &               19.28 &     1.4242 &          0.4792 &       145.279  &         4.5679 & WIS87           \\
 1987-Feb-02 & 06:54:54.432 & 2:07:55.373 &               13.64 &     1.4545 &          0.4896 &       144.818  &         6.7972 & WIS87           \\
 1991-Jul-07 & 20:22:11.222 & 0:52:58.109 &               41.98 &     1.2225 &          0.3046 &       314.682  &        14.7361 & VEL92           \\
 1991-Jul-17 & 20:39:38.390 & 0:16:11.395 &               42.25 &     1.1677 &          0.2189 &       322.928  &         9.3619 & VEL92           \\
 1996-Jul-14 & 21:49:35.616 & 1:42:18.720 &               40.29 &     1.1839 &          0.2343 &       319.231  &        11.525  & PRA98           \\
 1996-Jul-16 & 21:29:00.096 & 1:52:43.392 &               40.22 &     1.1731 &          0.218  &       320.915  &        10.2489 & PRA98           \\
 1996-Jul-19 & 20:01:42.182 & 1:20:49.978 &               40.26 &     1.1571 &          0.1949 &       323.577  &         8.0107 & PRA98           \\
 1996-Jul-19 & 21:50:23.136 & 1:49:38.496 &               40.26 &     1.1567 &          0.1943 &       323.649  &         7.9468 & PRA98           \\
 1996-Jul-21 & 22:21:37.152 & 1:33:29.952 &               40.48 &     1.1458 &          0.1794 &       325.621  &         6.0984 & PRA98           \\
 1996-Jul-26 & 22:47:04.704 & 2:05:27.168 &               42.47 &     1.119  &          0.1468 &       331.076  &         0.0887 & PRA98           \\
 1997-Feb-04 & 18:57:38.333 & 1:54:03.802 &                9.95 &     1.4576 &          0.4824 &       142.691  &         7.245  & PRA98           \\
 1997-Feb-04 & 22:49:04.195 & 3:26:27.427 &                9.86 &     1.4584 &          0.483  &       142.676  &         7.3004 & DUR12           \\
 1997-Mar-07 & 20:47:01.536 & 2:26:37.248 &               23.66 &     1.5993 &          0.7078 &       142.513  &        14.3409 & PRA98           \\
 2001-Jun-24 & 20:37:55.402 & 0:35:59.309 &               41.33 &     1.295  &          0.4229 &       305.437  &        18.6822 & DUR12           \\
 2002-Feb-16 & 17:31:59.002 & 2:59:19.536 &               11.69 &     1.5136 &          0.543  &       142.021  &        10.7815 & DUR12           \\
 2006-Jun-28 & 20:10:09.696 & 3:48:58.176 &               42.34 &     1.2695 &          0.3884 &       309.085  &        17.2568 & DUR12           \\
 2006-Jun-29 & 21:39:57.946 & 0:02:10.032 &               42.35 &     1.2637 &          0.3781 &       309.779  &        16.9759 & DUR12           \\
 2006-Jun-30 & 21:08:41.683 & 0:00:57.802 &               42.36 &     1.2583 &          0.3686 &       310.427  &        16.704  & DUR12           \\
 2006-Jul-25 & 21:54:52.531 & 3:58:06.038 &               46.2  &     1.1218 &          0.1632 &       332.464  &         0.7458 & DUR12           \\
 2006-Jul-25 & 22:22:34.608 & 3:21:52.733 &               46.21 &     1.1217 &          0.163  &       332.487  &         0.7206 & DUR12           \\
 2007-Feb-10 & 04:23:01.824 & 0:17:19.248 &                9.43 &     1.4856 &          0.5093 &       143.445  &         9.2355 & DUR12           \\
 2007-Feb-12 & 04:50:24.893 & 8:31:49.411 &                9.6  &     1.4952 &          0.5192 &       143.284  &         9.8281 & DUR12           \\
 2007-Feb-17 & 03:24:04.867 & 8:03:41.501 &               11.6  &     1.5184 &          0.5477 &       142.973  &        11.1469 & DUR12           \\
 2007-Feb-17 & 08:11:48.106 & 2:46:47.914 &               11.71 &     1.5194 &          0.549  &       142.963  &        11.195  & DUR12           \\
 2007-Feb-18 & 19:59:42.432 & 2:33:38.880 &               12.59 &     1.5262 &          0.5588 &       142.904  &        11.5563 & DUR12           \\
 2009-Mar-21 & 21:50:27.110 & 2:47:46.723 &               24.16 &     1.9018 &          1.1131 &       210.032  &        28.0154 & DUR12           \\
 2009-Mar-28 & 20:46:23.434 & 2:24:54.432 &               22.61 &     1.9026 &          1.0768 &       210.217  &        28.9456 & DUR12           \\
 2009-Apr-15 & 18:29:56.429 & 0:44:48.682 &               20.28 &     1.8983 &          1.0229 &       209.708  &        30.5989 & DUR12           \\
 2009-May-17 & 19:20:18.269 & 0:55:09.898 &               25.64 &     1.8673 &          1.0752 &       208.826  &        30.0859 & DUR12           \\
 2009-May-18 & 17:34:34.522 & 3:12:02.966 &               25.89 &     1.866  &          1.0791 &       208.863  &        30.0127 & DUR12           \\
 2009-May-24 & 17:54:09.043 & 9:46:08.717 &               27.48 &     1.8567 &          1.1072 &       209.226  &        29.4834 & DUR12           \\
 2011-May-31 & 21:30:36.346 & 0:26:52.742 &               42.35 &     1.4105 &          0.6883 &       294.664  &        20.7791 & DUR12           \\
 2011-Jun-03 & 22:52:19.200 & 0:44:40.560 &               42.57 &     1.3948 &          0.6545 &       296.31   &        20.5043 & DUR12           \\
 2011-Jun-03 & 23:53:48.480 & 1:31:11.194 &               42.57 &     1.3946 &          0.654  &       296.334  &        20.5003 & DUR12           \\
 2011-Jun-04 & 21:26:44.275 & 0:07:24.528 &               42.64 &     1.3899 &          0.6442 &       296.823  &        20.412  & DUR12           \\
 2011-Jun-04 & 22:52:47.971 & 0:38:47.443 &               42.64 &     1.3896 &          0.6435 &       296.856  &        20.4061 & DUR12           \\
 2011-Jun-05 & 21:46:59.232 & 3:08:55.565 &               42.71 &     1.3846 &          0.6331 &       297.38   &        20.3082 & DUR12           \\
 2011-Jun-05 & 22:43:32.678 & 0:40:05.462 &               42.71 &     1.3844 &          0.6326 &       297.402  &        20.3042 & DUR12           \\
 2011-Jun-08 & 21:34:43.795 & 0:36:51.235 &               42.92 &     1.3689 &          0.6006 &       299.045  &        19.9744 & DUR12           \\
 2011-Jun-09 & 21:18:03.283 & 0:29:00.701 &               42.99 &     1.3637 &          0.59   &       299.604  &        19.8543 & DUR12           \\
 2011-Jun-10 & 21:26:35.894 & 3:12:57.571 &               43.06 &     1.3584 &          0.5792 &       300.177  &        19.7269 & DUR12           \\
 2011-Jun-24 & 22:00:21.110 & 1:03:30.931 &               44.05 &     1.2828 &          0.4345 &       308.779  &        17.2486 & DUR12           \\
 2011-Jun-26 & 22:02:13.776 & 0:16:33.600 &               44.21 &     1.2719 &          0.4147 &       310.127  &        16.7548 & DUR12           \\
 2011-Jun-27 & 22:07:33.110 & 0:58:19.891 &               44.29 &     1.2664 &          0.4049 &       310.817  &        16.4897 & DUR12           \\
 2011-Jul-02 & 04:34:20.669 & 9:10:07.997 &               44.67 &     1.2429 &          0.364  &       313.871  &        15.2133 & DUR12           \\
 2011-Jul-07 & 20:50:11.530 & 1:54:38.304 &               45.33 &     1.2117 &          0.3122 &       318.31   &        13.0347 & DUR12           \\
 2011-Jul-12 & 19:42:51.898 & 0:04:48.317 &               46.2  &     1.1845 &          0.2698 &       322.632  &        10.5143 & DUR12           \\
 2011-Jul-22 & 19:46:34.378 & 3:22:37.315 &               50.33 &     1.1304 &          0.1969 &       333.161  &         2.4984 & DUR12           \\
 2011-Jul-22 & 22:17:38.083 & 2:36:15.149 &               50.41 &     1.1298 &          0.1963 &       333.288  &         2.3847 & DUR12           \\
 2011-Jul-22 & 23:12:22.579 & 1:42:24.941 &               50.44 &     1.1296 &          0.196  &       333.334  &         2.3436 & DUR12           \\
 2011-Jul-24 & 21:18:31.450 & 0:54:38.362 &               51.96 &     1.1194 &          0.185  &       335.693  &         0.1338 & DUR12           \\
 2011-Jul-26 & 21:23:41.453 & 0:48:16.301 &               53.92 &     1.1088 &          0.175  &       338.268  &        -2.4533 & DUR12           \\
 2011-Jul-27 & 05:39:39.082 & 0:36:12.730 &               54.29 &     1.107  &          0.1733 &       338.715  &        -2.9253 & DUR12           \\
 2011-Aug-13 & 06:51:40.464 & 0:15:29.261 &               81.93 &     1.023  &          0.1704 &         1.786  &       -31.6103 & DUR12           \\
 2011-Aug-14 & 06:41:30.221 & 0:14:37.939 &               83.28 &     1.0185 &          0.1745 &         3.0365 &       -33.0291 & DUR12           \\
 2011-Oct-25 & 06:40:55.056 & 8:45:26.842 &               76.62 &     0.9527 &          0.581  &       100.483  &       -34.5387 & DUR12           \\
 2011-Oct-26 & 06:45:58.579 & 8:42:16.502 &               76.21 &     0.956  &          0.5839 &       101.508  &       -34.0919 & DUR12           \\
 2011-Oct-27 & 06:43:30.317 & 8:48:34.070 &               75.79 &     0.9593 &          0.5866 &       102.514  &       -33.6468 & DUR12           \\
\hline

\end{tabular}
\end{table*}


\begin{table*}
\caption{Continuation of table \ref{tab:Eger archive}}
\begin{tabular}{lllrrrrrl}
\hline
 Date        & UT (start)   & UT (end)    &   $\alpha [^\circ]$ &   $r$ [au] &   $\Delta$ [au] &   PABLon [deg] &   PABLat [deg] & Reference   \\
 \hline
  2011-Oct-29 & 06:36:12.787 & 8:46:36.566 &               74.97 &     0.9662 &          0.5916 &       104.486  &       -32.7562 & DUR12           \\
 2011-Oct-30 & 06:39:26.582 & 8:06:33.437 &               74.56 &     0.9698 &          0.5939 &       105.457  &       -32.3085 & DUR12           \\
 2011-Nov-01 & 06:48:30.384 & 8:38:00.413 &               73.74 &     0.9772 &          0.5982 &       107.361  &       -31.4125 & DUR12           \\
 2011-Nov-02 & 06:39:03.427 & 8:42:33.350 &               73.34 &     0.981  &          0.6001 &       108.286  &       -30.9691 & DUR12           \\
 2011-Nov-03 & 06:13:04.598 & 8:41:28.464 &               72.94 &     0.9849 &          0.6019 &       109.188  &       -30.5313 & DUR12           \\
 2011-Dec-09 & 01:01:18.221 & 3:28:25.363 &               58.16 &     1.1592 &          0.5874 &       134.694  &       -15.2567 & DUR12           \\
 2011-Dec-29 & 21:15:27.331 & 2:57:07.891 &               46.15 &     1.2735 &          0.5303 &       143.199  &        -6.737  & DUR12           \\
 2012-Jan-30 & 03:06:49.363 & 1:10:11.165 &               17.79 &     1.4384 &          0.488  &       146.03   &         5.9592 & DUR12           \\
 2014-Apr-20 & 04:22:33.571 & 1:36:08.006 &               20.39 &     1.894  &          1.0179 &       210.571  &        30.8605 & WAR17           \\
 2014-Apr-21 & 06:32:02.141 & 1:57:18.259 &               20.43 &     1.8933 &          1.0172 &       210.51   &        30.9005 & WAR17           \\
 2014-Apr-23 & 03:57:00.749 & 1:53:01.133 &               20.53 &     1.8919 &          1.0164 &       210.405  &        30.9576 & WAR17           \\
 2016-May-31 & 08:13:12.518 & 1:24:47.261 &               43.56 &     1.3971 &          0.6938 &       296.851  &        19.899  & WAR17           \\
 2016-Jun-01 & 08:14:37.709 & 1:11:25.296 &               43.66 &     1.3919 &          0.6827 &       297.42   &        19.7959 & WAR17           \\
 2016-Jun-02 & 08:07:52.234 & 0:56:13.085 &               43.77 &     1.3867 &          0.6717 &       297.99   &        19.6891 & WAR17           \\
 2016-Jun-03 & 08:05:50.669 & 1:20:24.691 &               43.88 &     1.3815 &          0.6607 &       298.566  &        19.5775 & WAR17           \\
 2016-Jun-04 & 08:04:31.786 & 1:16:02.122 &               43.99 &     1.3763 &          0.6498 &       299.147  &        19.4612 & WAR17           \\
 2017-Feb-05 & 19:31:47.136 & 0:55:15.658 &               13.65 &     1.4864 &          0.5228 &       147.534  &         9.2321 & DUR18           \\
 2017-Feb-15 & 23:59:19.046 & 4:56:09.024 &               11.36 &     1.534  &          0.5633 &       146.666  &        12.1105 & DUR18           \\
\hline
\end{tabular}
\end{table*}

\begin{table*}
\caption{Archival observations for (12711) Tukmit. The information includes the date, the starting and end time (UT) of the observations, the phase angle ($\alpha$), the heliocentric ($r$) and geocentric ($\Delta$) distances, phase angle bisector longitude (PABLon) and latitude (PABLat) of the asteroid at the time of observation. \textbf{References:} WS22: \protect\cite{2022MPBu...49...83W}.} 
\begin{tabular}{lllrrrrrl}
\hline
 Date        & UT (start)   & UT (end)    &   $\alpha [^\circ]$ &   $r$ [au] &   $\Delta$ [au] &   PABLon [deg] &   PABLat [deg] & Reference   \\
\hline
 2021-Nov-29 & 08:24:37.325 & 3:18:03.658 &               39.6  &     1.4849 &          0.8661 &        118.675 &        17.9726 & WS22           \\
 2021-Nov-30 & 08:25:33.830 & 3:27:22.234 &               39.42 &     1.4836 &          0.8541 &        118.936 &        17.6979 & WS22           \\
\hline
\end{tabular}
\label{tab:Tukmit archive}
\end{table*}

\begin{table*}
\caption{Archival observations for (161989) Cacus. The information includes the date, the starting and end time (UT) of the observations, the phase angle ($\alpha$), the heliocentric ($r$) and geocentric ($\Delta$) distances, phase angle bisector longitude (PABLon) and latitude (PABLat) of the asteroid at the time of observation. \textbf{References:} SCH79: \protect\cite{1979A&AS...37..483S}; DEG78: \protect\cite{1978IAUC.3193....1D}; KOE14: \protect\cite{2014MPBu...41..286K}; DUR18: \protect\cite{2018A&A...609A..86D}} 
\begin{tabular}{lllrrrrrl}
\hline
 Date        & UT (start)   & UT (end)    &   $\alpha [^\circ]$ &   $r$ [au] &   $\Delta$ [au] &   PABLon [deg] &   PABLat [deg] & Reference   \\
\hline
 1978-Mar-01 & 02:10:20.064 & 8:04:20.554 &                8.44 &     1.1316 &          0.1425 &       156.682  &        -3.9248 & SCH79           \\
 1978-Mar-08 & 03:58:30.691 & 9:03:15.091 &               25.62 &     1.1069 &          0.1284 &       153.717  &         8.1938 & DEG78           \\
 2003-Feb-18 & 00:32:52.253 & 6:25:12.432 &               28.51 &     1.1863 &          0.2325 &       146.066  &       -20.5229 & DUR18           \\
 2003-Mar-05 & 18:11:19.622 & 9:08:36.701 &               35.15 &     1.1341 &          0.1807 &       141.547  &        -4.6256 & DUR18           \\
 2003-Mar-25 & 18:49:27.408 & 3:21:20.160 &               67.17 &     1.0636 &          0.2294 &       143.1    &        21.3483 & DUR18           \\
 2003-Apr-01 & 19:15:52.243 & 0:08:22.301 &               74.1  &     1.0388 &          0.2618 &       146.707  &        28.8395 & DUR18           \\
 2003-Apr-04 & 20:01:34.234 & 0:58:15.715 &               76.45 &     1.0282 &          0.2763 &       148.628  &        31.8365 & DUR18           \\
 2009-Feb-19 & 09:04:52.550 & 2:29:30.422 &               50.96 &     1.1211 &          0.2379 &       186.977  &         0.3458 & KOE14           \\
 2014-Dec-21 & 05:34:44.371 & 8:31:08.112 &               51.19 &     1.2536 &          0.9041 &       159.459  &       -17.1859 & DUR18           \\
 2015-Feb-17 & 06:55:48.605 & 9:04:36.307 &               67.61 &     1.0667 &          0.4663 &       207.04   &        11.6119 & DUR18           \\
 2015-Feb-17 & 08:16:10.675 & 8:31:27.379 &               67.63 &     1.0665 &          0.4661 &       207.097  &        11.6582 & DUR18           \\
 2015-Oct-09 & 07:26:34.714 & 8:55:20.410 &               50.18 &     1.3003 &          0.8079 &        72.094  &       -35.7728 & DUR18           \\
 2015-Oct-13 & 07:30:26.179 & 9:07:28.762 &               49.64 &     1.3081 &          0.7994 &        74.1272 &       -36.8006 & DUR18           \\
 2015-Nov-05 & 05:58:17.011 & 8:49:43.018 &               46.57 &     1.3433 &          0.7449 &        84.312  &       -41.7277 & DUR18           \\
 2015-Dec-08 & 05:14:32.352 & 8:38:07.757 &               42.53 &     1.3632 &          0.6561 &        93.524  &       -45.3206 & DUR18           \\
 2015-Dec-15 & 04:48:22.032 & 8:25:47.741 &               41.86 &     1.3627 &          0.6381 &        94.5876 &       -45.3522 & DUR18           \\
 2016-Feb-04 & 03:24:48.499 & 5:03:14.803 &               44.61 &     1.3093 &          0.5769 &       100.413  &       -33.8912 & DUR18           \\
 2016-Feb-12 & 00:23:39.811 & 1:57:33.005 &               46.51 &     1.2936 &          0.5847 &       102.37   &       -30.374  & DUR18           \\
 2016-Mar-09 & 23:53:28.608 & 3:32:30.739 &               53.87 &     1.2258 &          0.6441 &       112.676  &       -16.8207 & DUR18           \\
 2016-Oct-05 & 23:43:01.776 & 3:23:46.378 &               66.03 &     1.0888 &          0.3437 &       323.976  &       -17.9056 & DUR18           \\
 2016-Dec-22 & 00:37:07.046 & 3:09:58.061 &               48.43 &     1.3102 &          0.9512 &        23.6428 &       -32.6566 & DUR18           \\
 2016-Dec-31 & 01:00:04.435 & 3:47:11.587 &               47.42 &     1.3258 &          1.015  &        29.5956 &       -32.4289 & DUR18           \\
\hline
\end{tabular}
\label{tab:Cacus archive}
\end{table*}

\section{Statistical quality of pole solutions}

\begin{figure*}
    \centering
    \includegraphics[width=\linewidth,keepaspectratio]{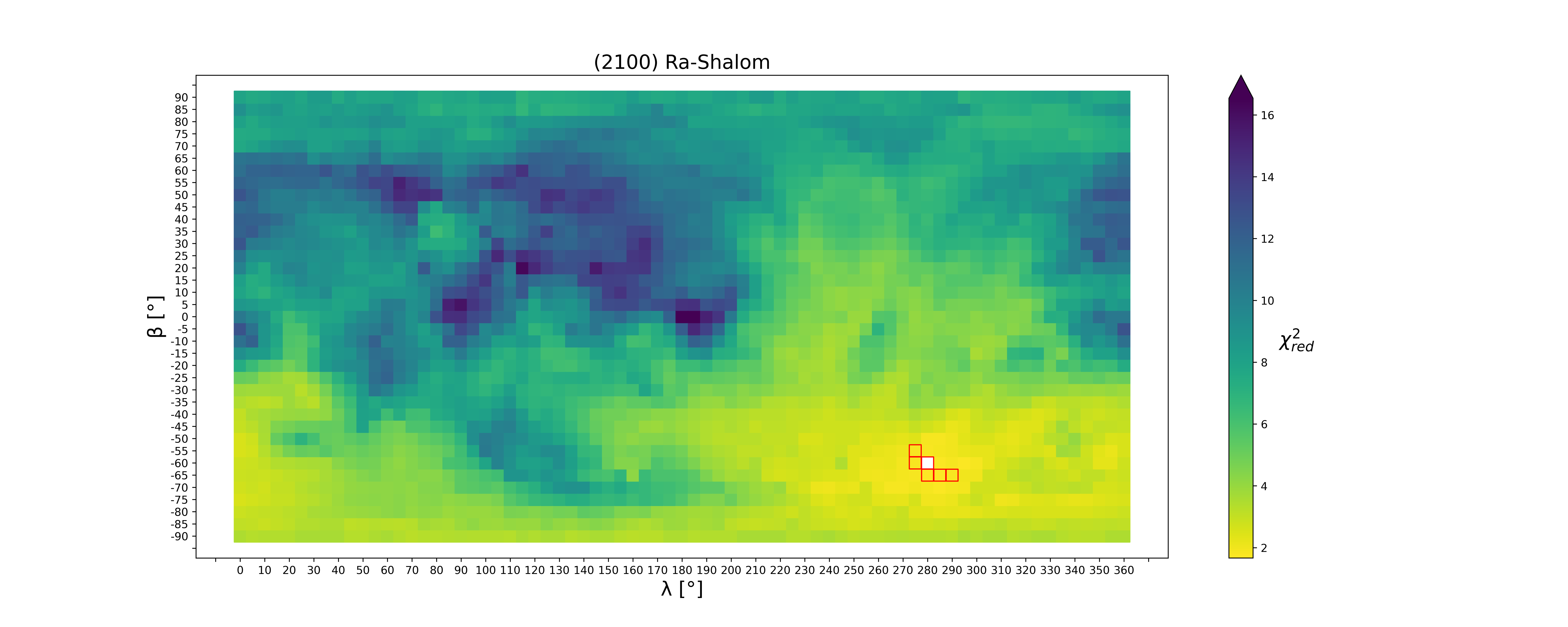}
    \caption{Statistical quality of (2100) Ra-Shalom pole solutions obtained with the constant period code. The solutions are shaded by its $\chi_\mathrm{red}^{2}$ value. The best solution obtained is shown as a white square ($\lambda=278 ^{\circ}, \beta=-60 ^{\circ}$) with $\chi_\mathrm{red}^{2}=1.66$ (normalized by the 4987 data points). The solutions within a margin of 5.7\% (3$\sigma$) are highlighted with a red border.}
    \label{fig:Ra-Shalom pole plot no yorp}
\end{figure*}

\begin{figure*}
    \centering
    \includegraphics[width=\linewidth,keepaspectratio]{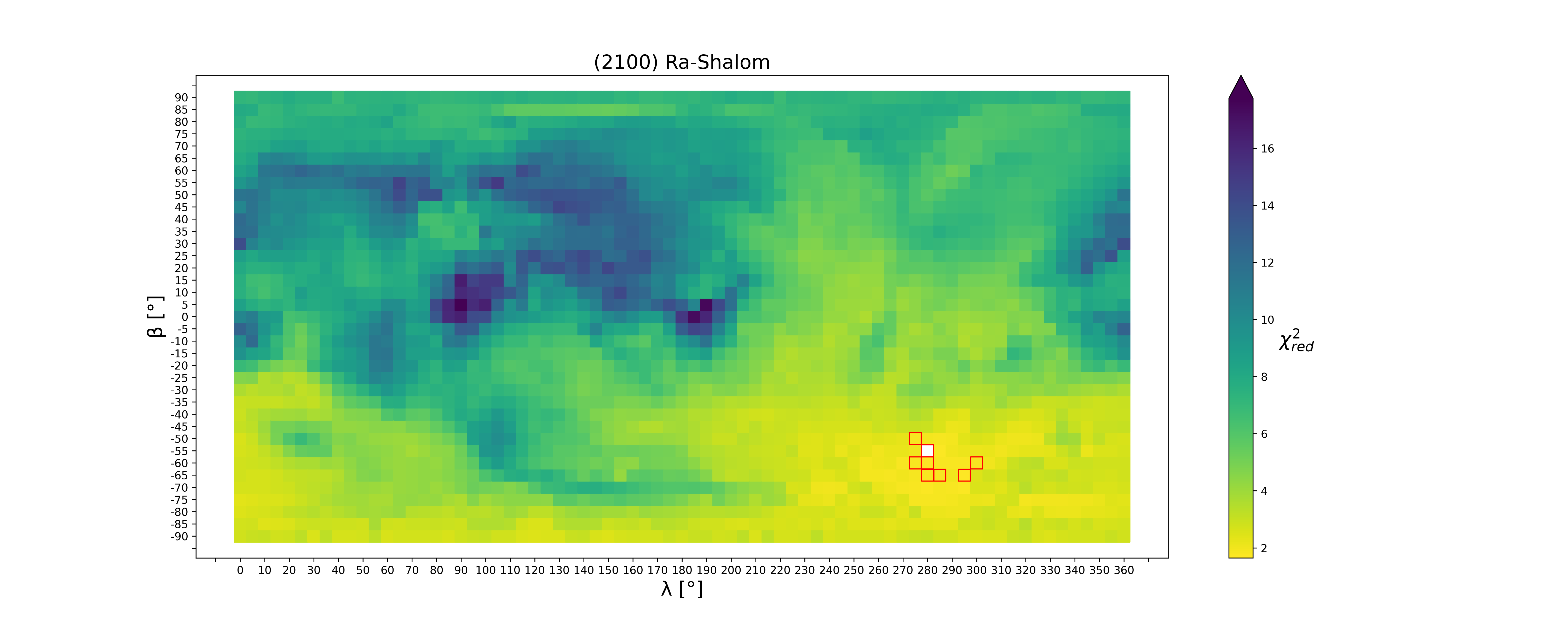}
    \caption{Statistical quality of (2100) Ra-Shalom pole solutions obtained with the linear increasing period code. The solutions are shaded by its $\chi_\mathrm{red}^{2}$ value. The best solution obtained is shown as a white square ($\lambda=283 ^{\circ}, \beta=-62 ^{\circ}$) with $\chi_\mathrm{red}^{2}=1.64$ (normalized by the 4987 data points). The solutions within a margin of 5.7\% (3$\sigma$) are highlighted with a red border.}
    \label{fig:Ra-Shalom pole plot}
\end{figure*}

\begin{figure*}
    \centering
    \includegraphics[width=\linewidth,keepaspectratio]{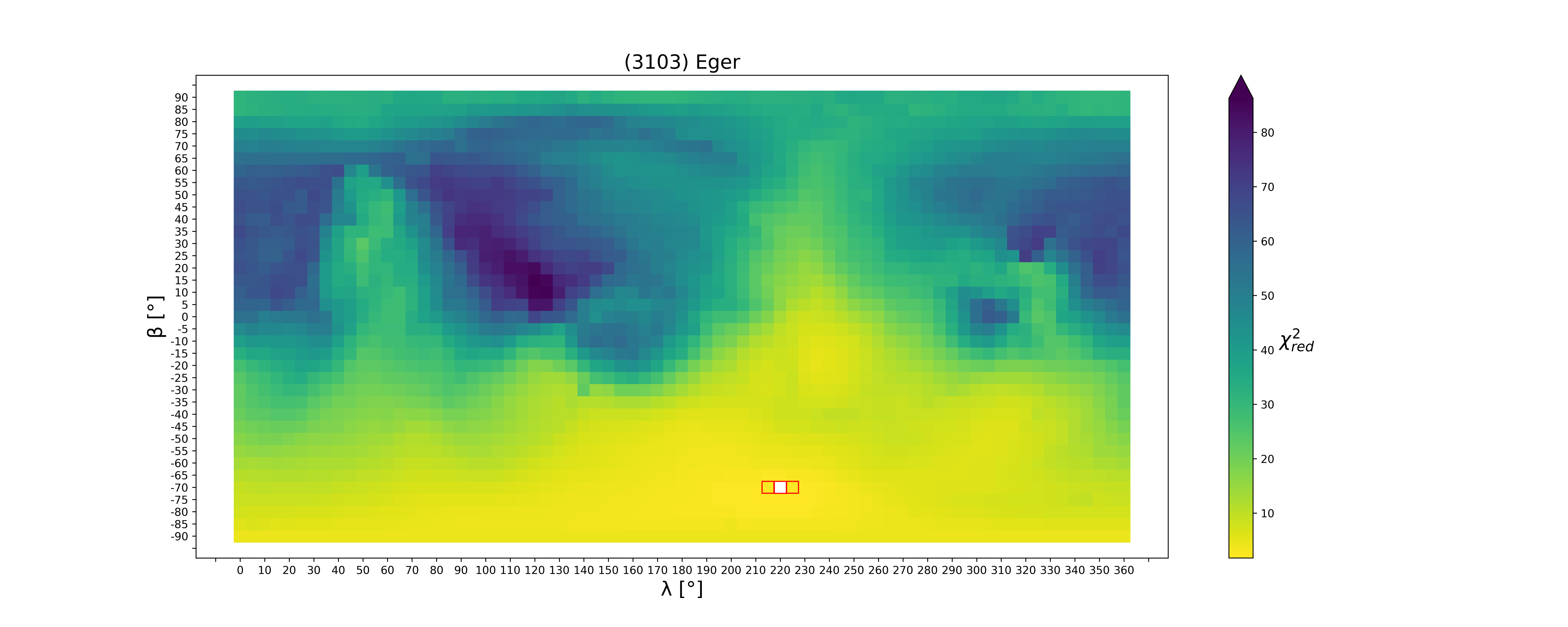}
    \caption{Statistical quality of (3103) Eger pole solutions obtained with the linear increasing period code. The solutions are shaded by its $\chi_\mathrm{red}^{2}$ value. The best solution obtained is shown as a white square ($\lambda=214 ^{\circ}, \beta=-71 ^{\circ}$) with a $\chi_\mathrm{red}^{2}=1.74$ (normalized by the 6034 data points), the solutions within a margin of 5.2\% (3$\sigma$) are highlighted with a red border.}
    \label{fig:Eger pole plot}
\end{figure*}

\begin{figure*}
    \centering
    \includegraphics[width=\linewidth,keepaspectratio]{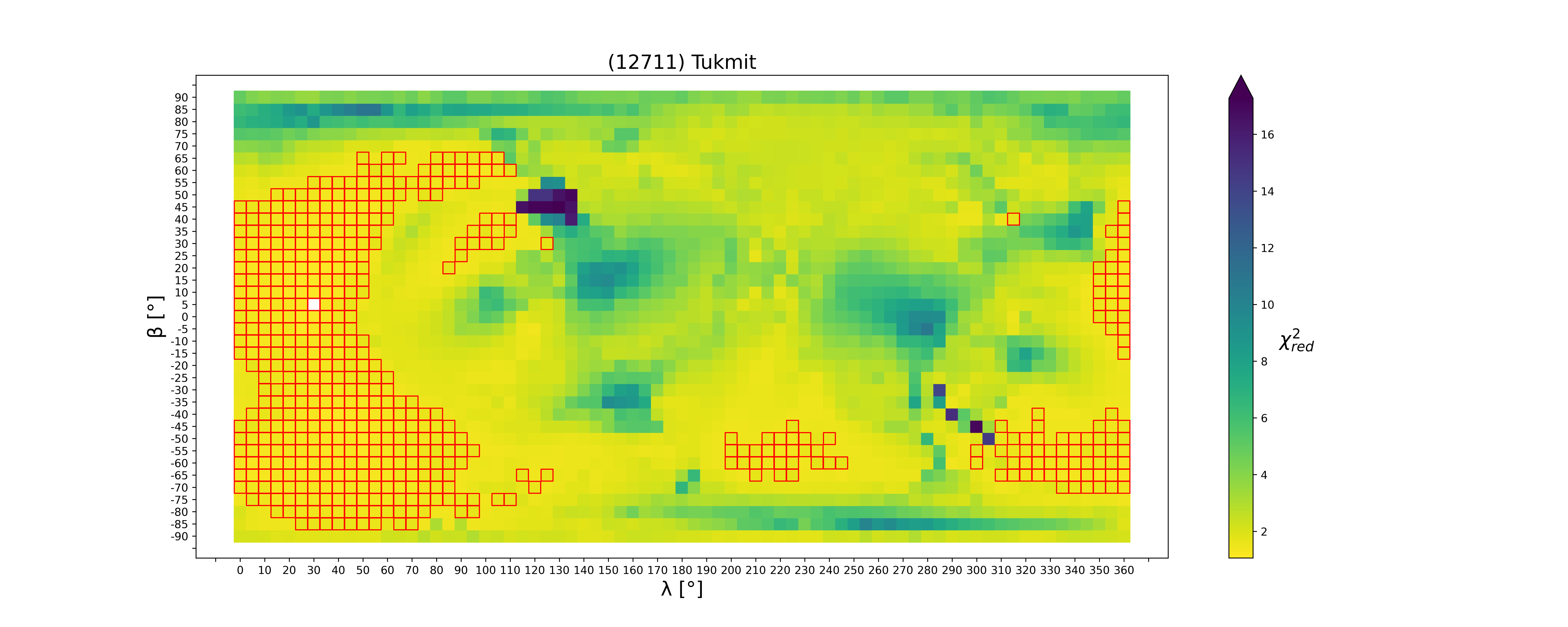}
    \caption{Statistical quality of (12711) Tukmit pole solutions obtained with the constant period code. The solutions are shaded by its $\chi_\mathrm{red}^{2}$ value. The best solution obtained is shown as a white square ($\lambda=27 ^{\circ}, \beta=11 ^{\circ}$) with a $\chi_\mathrm{red}^{2}=1.06$ (normalized by the 263 data points), the solutions within a margin of 25\% (3$\sigma$) are highlighted with a red border.}
    \label{fig:Tukmit pole plot}
\end{figure*}

\begin{figure*}
    \centering
    \includegraphics[width=\linewidth,keepaspectratio]{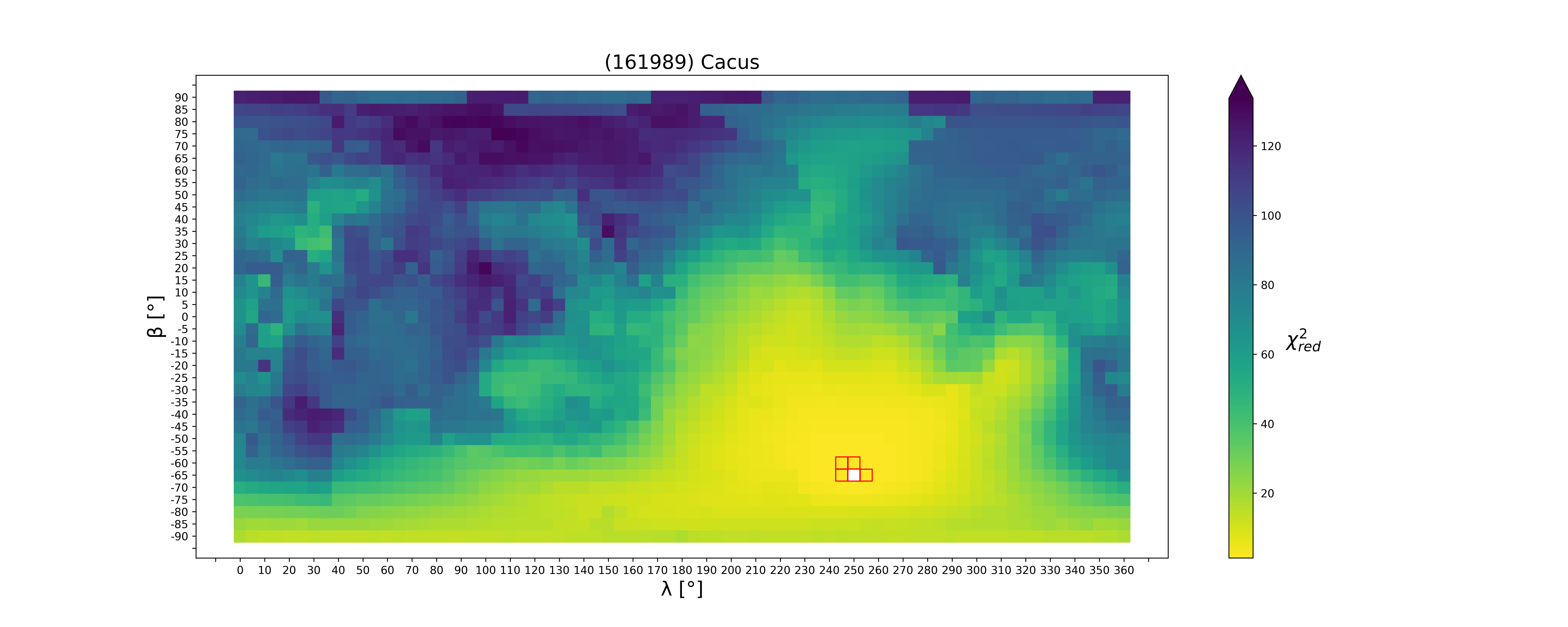}
    \caption{Statistical quality of (161989) Cacus pole solutions obtained with the linear increasing period code. The solutions are shaded by its $\chi_\mathrm{red}^{2}$ value. The best solution obtained is shown as a white square ($\lambda=251 ^{\circ}, \beta=-61 ^{\circ}$) with a $\chi_\mathrm{red}^{2}=1.31$ (normalized by the 1534 data points), the solutions within a margin of 10\% (3$\sigma$) are highlighted with a red border.}
    \label{fig:Cacus pole plot}
\end{figure*}

\section{Fits of models and data}
\label{sec: Fits of models and data}

\begin{figure*}
    \centering
    \begin{subfigure}{0.33\textwidth}
        \centering
        \includegraphics[width=\textwidth]{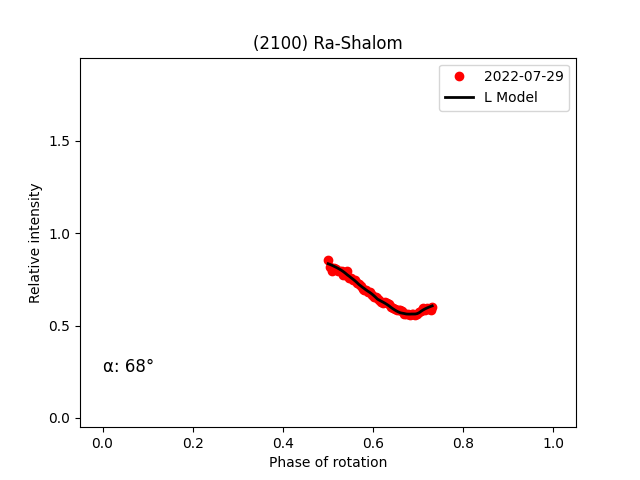}
    \end{subfigure}
    \begin{subfigure}{0.33\textwidth}
        \centering
        \includegraphics[width=\textwidth]{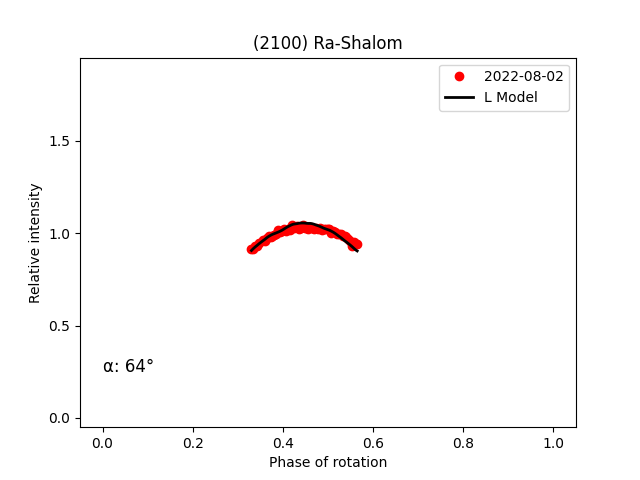}
    \end{subfigure}
    \begin{subfigure}{0.33\textwidth}
        \centering
        \includegraphics[width=\textwidth]{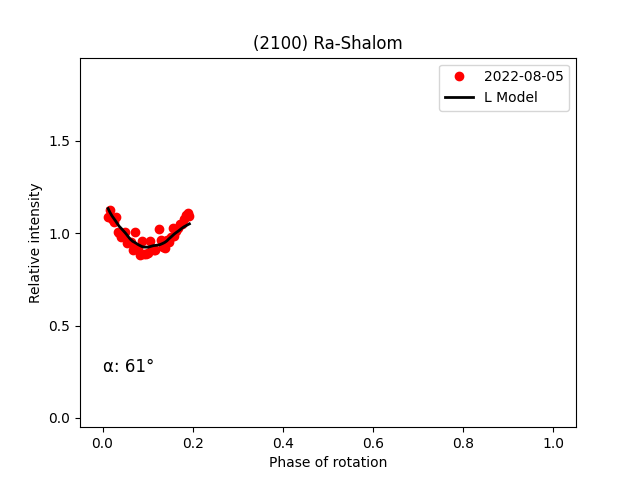}
    \end{subfigure}
    \begin{subfigure}{0.33\textwidth}
        \centering
        \includegraphics[width=\textwidth]{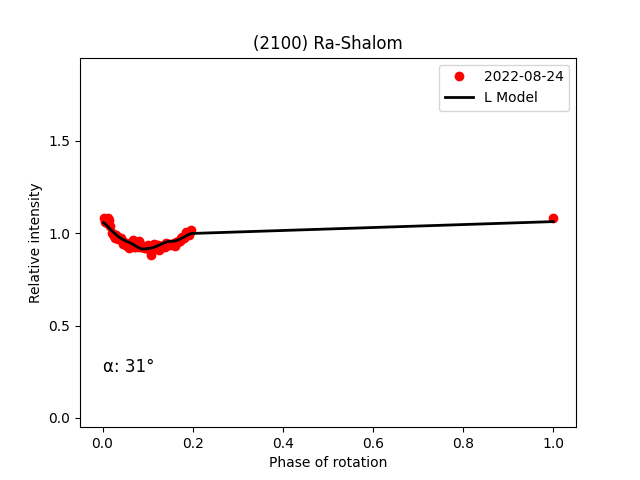}
    \end{subfigure}
    \begin{subfigure}{0.33\textwidth}
        \centering
        \includegraphics[width=\textwidth]{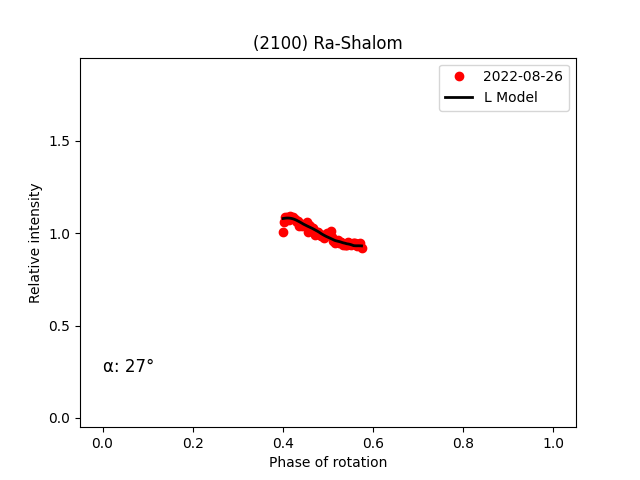}
    \end{subfigure}
    \begin{subfigure}{0.33\textwidth}
        \centering
        \includegraphics[width=\textwidth]{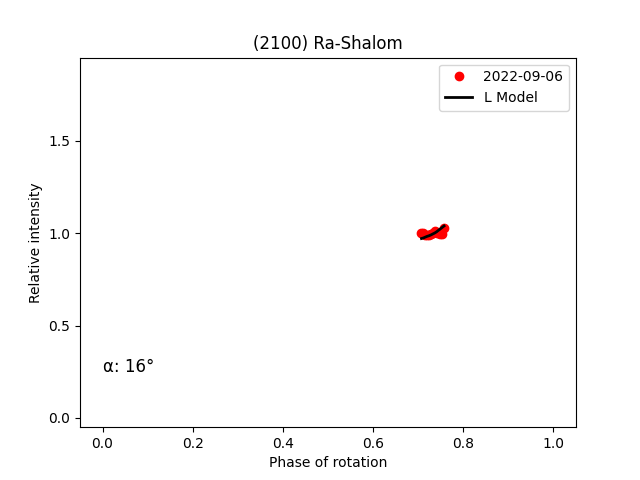}
    \end{subfigure}
    \begin{subfigure}{0.33\textwidth}
        \centering
        \includegraphics[width=\textwidth]{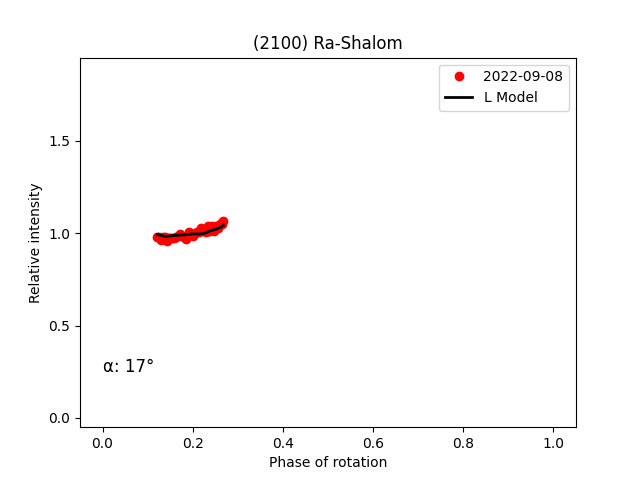}
    \end{subfigure}
    \caption{Fit between lightcurves from (2100) Ra-Shalom presented in this paper and the best-fitting linearly increasing period model (L Model). The data is plotted as dots for each observation, meanwhile the model is plotted as a solid black line. The geometry is described its solar phase angle $\alpha$.}
    \label{fig:IAC fit Ra-Shalom}
\end{figure*}

\begin{figure*}
    \centering
        \begin{subfigure}{0.33\textwidth}
        \centering
        \includegraphics[width=\textwidth]{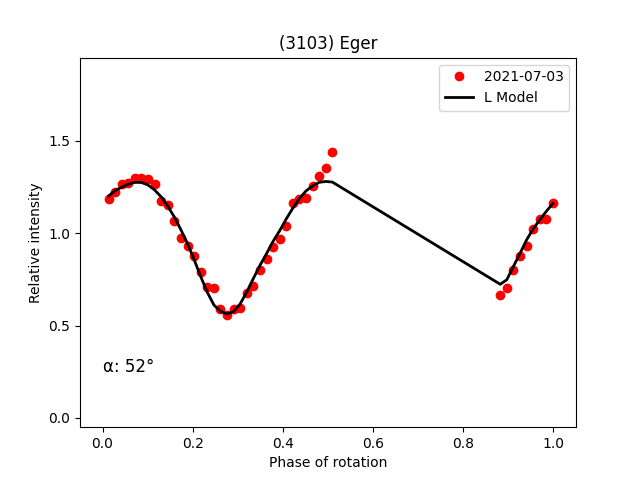}
    \end{subfigure}
    \begin{subfigure}{0.33\textwidth}
        \centering
        \includegraphics[width=\textwidth]{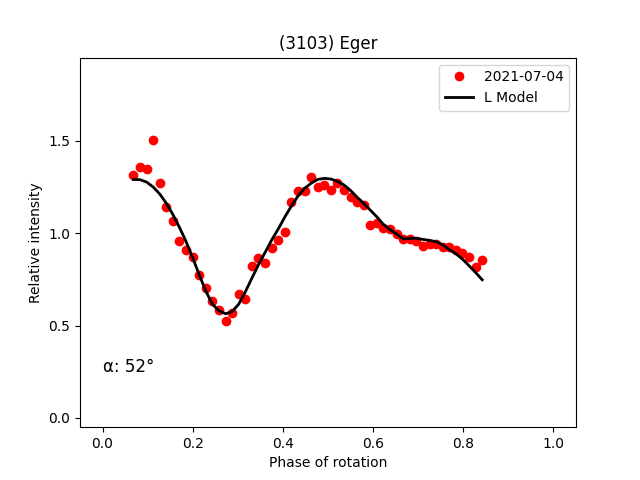}
    \end{subfigure}
    \begin{subfigure}{0.33\textwidth}
        \centering
        \includegraphics[width=\textwidth]{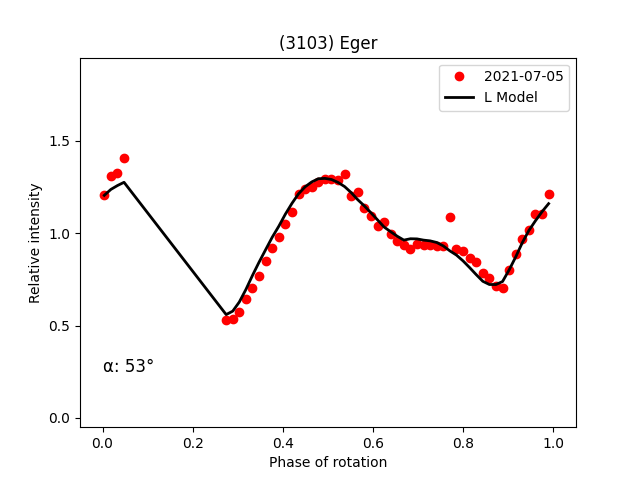}
    \end{subfigure}
    \begin{subfigure}{0.33\textwidth}
        \centering
        \includegraphics[width=\textwidth]{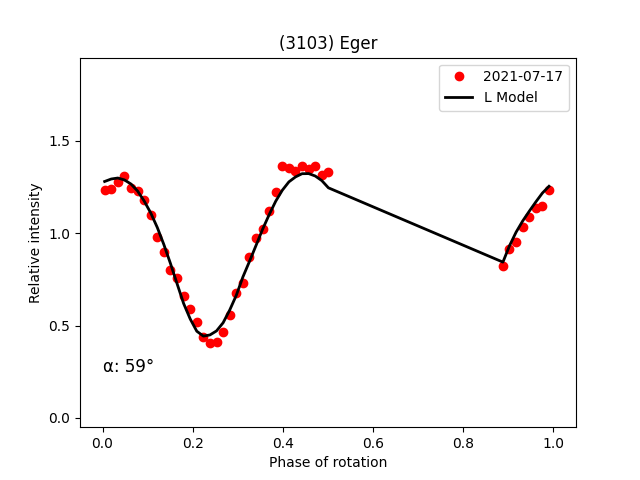}
    \end{subfigure}
    \begin{subfigure}{0.33\textwidth}
        \centering
        \includegraphics[width=\textwidth]{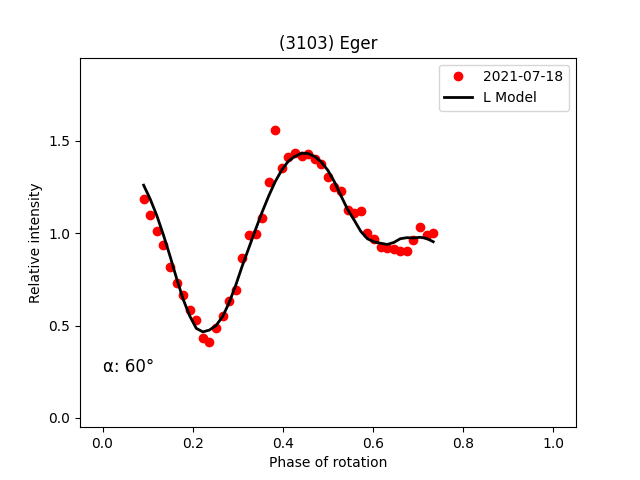}
    \end{subfigure}
    \begin{subfigure}{0.33\textwidth}
        \centering
        \includegraphics[width=\textwidth]{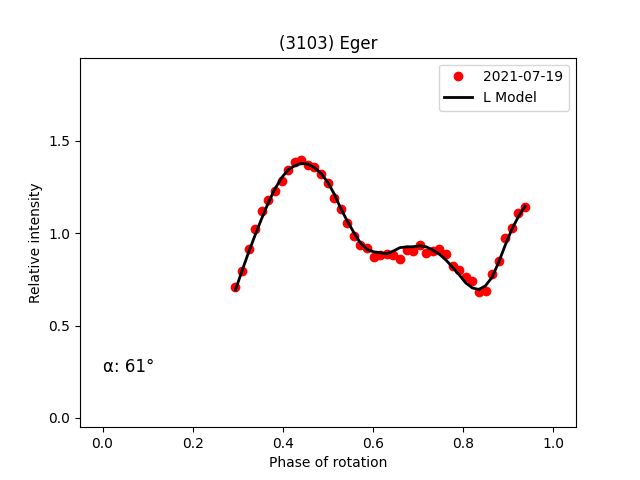}
    \end{subfigure}
    \begin{subfigure}{0.33\textwidth}
        \centering
        \includegraphics[width=\textwidth]{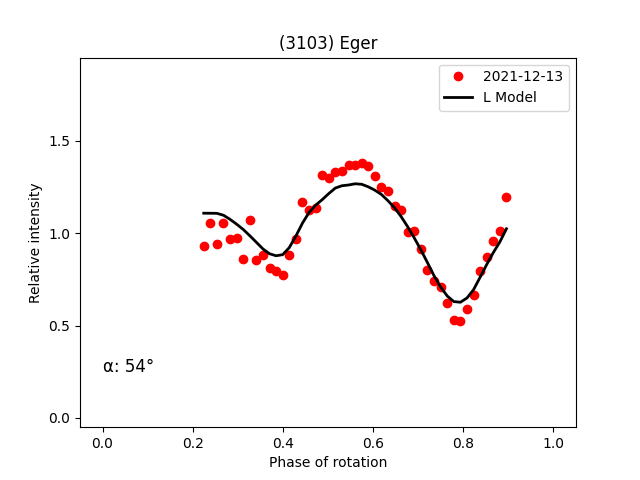}
    \end{subfigure}
    \begin{subfigure}{0.33\textwidth}
        \centering
        \includegraphics[width=\textwidth]{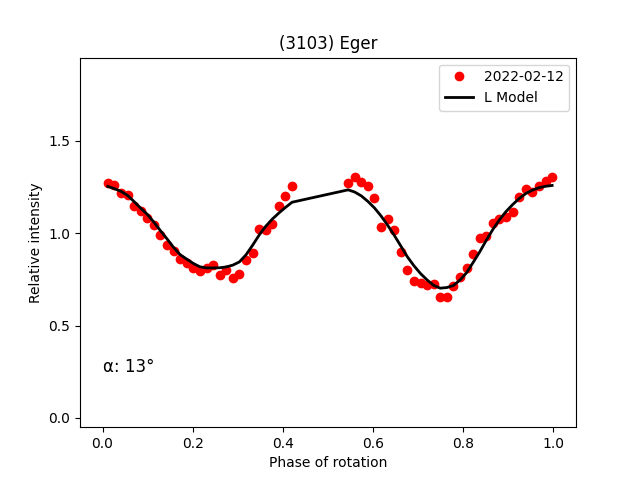}
    \end{subfigure}
    \begin{subfigure}{0.33\textwidth}
        \centering
        \includegraphics[width=\textwidth]{Figures/Eger/89_IAC_Eger.png}
    \end{subfigure}
    \begin{subfigure}{0.33\textwidth}
        \centering
        \includegraphics[width=\textwidth]{Figures/Eger/90_IAC_Eger.png}
    \end{subfigure}
    \caption{Fit between lightcurves from (3103) Eger presented in this paper and the best-fitting linearly increasing period model (L Model). The data is plotted as red dots for each observation, meanwhile the model is plotted as a solid black line. The geometry is described its solar phase angle $\alpha$.}
    \label{fig:IAC fit Eger}
\end{figure*}

\begin{figure*}
    \centering
    \begin{subfigure}{0.49\textwidth}
        \centering
        \includegraphics[width=\textwidth]{Figures/Cacus/23_IAC_Cacus.png}
    \end{subfigure}
    \begin{subfigure}{0.49\textwidth}
        \centering
        \includegraphics[width=\textwidth]{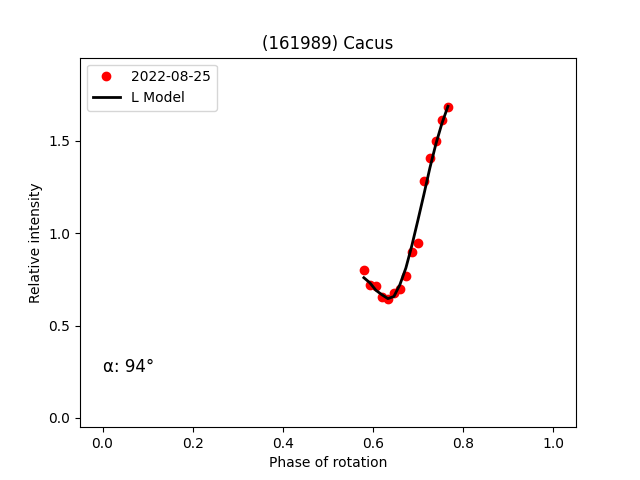}
    \end{subfigure}
    \begin{subfigure}{0.49\textwidth}
        \centering
        \includegraphics[width=\textwidth]{Figures/Cacus/25_IAC_Cacus.png}
    \end{subfigure}
    \caption{Fit between lightcurves from (161989) Cacus presented in this paper and the best-fitting linearly increasing period model (L Model). The data is plotted as red dots for each observation, meanwhile the model is plotted as a solid black line. The geometry is described its solar phase angle $\alpha$.}
    \label{fig:IAC fit Cacus}
\end{figure*}


\bsp	
\label{lastpage}
\end{document}